\documentclass[sigconf]{acmart}

\AtBeginDocument{%
  \providecommand\BibTeX{{%
    \normalfont B\kern-0.5em{\scshape i\kern-0.25em b}\kern-0.8em\TeX}}}

\copyrightyear{2026}
\acmYear{2026}
\setcopyright{cc}
\setcctype{by}
\acmConference[NordiCHI '26]{Proceedings of the 14th Nordic Conference on Human-
Computer Interaction}{October 03--07, 2026}{Vaasa, Finland}
\acmBooktitle{Proceedings of the 14th Nordic Conference on Human-Computer
Interaction (NordiCHI '26), October 03--07, 2026, Vaasa, Finland}
\acmDOI{10.1145/3829807.3829837}
\acmISBN{979-8-4007-2373-5/2026/10}

\definecolor{mossygreen}{RGB}{85,107,47}
\definecolor{lightmossygreen}{RGB}{200,220,180}

\usepackage{multicol}
\usepackage{multirow}
\usepackage{tabularx}
\usepackage{soul}
\usepackage{wasysym}
\usepackage{pifont} 
\usepackage{cancel}
\usepackage{graphicx}
\usepackage{subcaption}
\usepackage{float}
\usepackage{balance}

\usepackage{placeins}

\begin{document}

\title[Disentangling Innovation Practices]{Disentangling Innovation Practices in Automation-Adopting Organizations: a Co-Performance Perspective}


\author{Garoa Gomez-Beldarrain}
\orcid{0009-0002-0141-6553}
\affiliation{
  \institution{Delft University of Technology}
  \city{Delft}
  \country{Netherlands}
}
\email{G.GomezBeldarrain@tudelft.nl}

\author{Kars Alfrink}
\orcid{0000-0001-7562-019X}
\affiliation{
  \institution{Delft University of Technology}
  \city{Delft}
  \country{Netherlands}
}
\email{C.P.Alfrink@tudelft.nl}

\author{Euiyoung Kim}
\orcid{0000-0003-2992-7718}
\affiliation{
  \institution{Delft University of Technology}
  \city{Delft}
  \country{Netherlands}
}
\email{E.Y.Kim@tudelft.nl}

\author{Elisa Giaccardi}
\orcid{0000-0002-3292-2531}
\affiliation{
  \institution{Politecnico di Milano}
  \city{Milano}
  \country{Italy}
}
\email{elisa.giaccardi@polimi.it}

\author{Alessandro Bozzon}
\orcid{0000-0002-3300-2913}
\affiliation{
  \institution{Delft University of Technology}
  \city{Delft}
  \country{Netherlands}
}
\email{A.Bozzon@tudelft.nl}

\author{Himanshu Verma}
\orcid{0000-0002-2494-1556}
\affiliation{
  \institution{Delft University of Technology}
  \city{Delft}
  \country{Netherlands}
}
\email{H.Verma@tudelft.nl}







\renewcommand{\shortauthors}{Gomez-Beldarrain et al.}

\begin{abstract} 




As organizations increasingly adopt automation, 
innovation practitioners are responsible for selecting, adapting, testing, and implementing externally sourced innovations.
However, little is known about how these upstream practices shape worker-automation arrangements, limiting our ability to intervene in innovation practice to address automation adoption challenges.
  To disentangle this relationship, 
  we interviewed nine innovation practitioners at a major European airport pursuing long-term autonomous operations and analyzed their practices through a co-performance lens. We synthesize five co-performance design principles and examine where current practices align or conflict. Our findings reveal tensions: innovation practitioners prioritize full-automation arrangements while postponing human considerations; contextual constraints shape solutions, but openness to reconfiguration remains limited; and co-learning rarely extends beyond pilot phases. 
  These insights provide HCI research and practice with guidance for reframing the conceptualization of automation, particularly by encouraging earlier consideration of human roles, promoting iterative visions, and recognizing workers as co-designers throughout innovation pipelines.


\end{abstract}

\color{black}




\maketitle

\section{Introduction}\label{sec:introduction}

With the advent of Artificial Intelligence (AI), organizations are increasingly adopting automated operations\footnote{In this work, we set the scope in automation, understood as systems that partially or fully shift manual work to machines within organizations. 
AI is an enabling technology driving automation, with applications such as robotics and autonomous driving.
Given the significant overlap between the two, with many concepts studied in AI research (e.g., worker replacement) having implications for automation, we draw on AI research to inform our research. 
We acknowledge, however, that the two concepts do not fully overlap: not all automation is based on AI, and not all AI leads to automated applications.}
to replace manual work processes traditionally performed by human labor~\cite{IVANOV2020, Eisser2020, Acemoglu2019, Moradi2025}. 
Promises of productivity and efficiency~\cite{Eisser2020, IVANOV2020, Wang2019, Kallbacker2025}, 
make automation innovation---its selection, development, and incorporation into broader organizational and procedural contexts---an opportunity to address current capacity issues \cite{Gamkrelidze2024}, 
worker shortages \cite{Wang2019}, and working conditions \cite{Roto2024, Unhelkar2014, Baldauf2021ws}. 
However, automation innovation involves challenging, complex, and often prolonged processes \cite{GYLDENKAERNE2024, Zajac2023}, and many projects ultimately see poor adoption in practice~\cite{GomezBeldarrain2025, Cabitza2020, Russo2024, Molin2024}. 
Across fields such as healthcare \cite{GYLDENKAERNE2024, Gamkrelidze2024}, aviation \cite{Gomez-Beldarrain2024}, public transportation \cite{Akridge2024}, and other essential sectors \cite{Fox2023}, automation prototypes and pilots frequently fail to scale \cite{GYLDENKAERNE2024} or integrate smoothly into daily operations \cite{Fox2023}.

One key factor underlying these failures is how appropriate worker\footnote{In this paper, we use the term `worker' to refer specifically to frontline personnel conducting physical ground operations, who would be affected by automation in their daily work. We use `innovation practitioner' to refer to professionals involved in selecting, adapting, testing, and scaling automation within organizations, typically as part of innovation teams.}-automation arrangements are conceptualized within organizations, as extensively documented in prior HCI research. 
Automation is 
often framed as a quick  fix (cf.~\cite{Mlynar2022}) for organizational challenges~\cite{GomezBeldarrain2025, Riek2025, Gamkrelidze2024}, relying on myths \cite{Bradshaw2013} and reductionist views of worker practices~\cite{Akridge2024, Breuer2023}.
As a result,  initial investments in automation projects tend to focus solely on technology development, while underlying organizational issues \cite{Gamkrelidze2024} 
and involved workers are treated as an  afterthought~\cite{GomezBeldarrain2025}. 
These dynamics lead to 
 poorly conceived automation implementations and
 well-documented \textit{ironies of automation} \cite{bainbridge1983}, including misplaced responsibilities, skill degradation, and additional burdens on workers who must supervise, step in, or remain in the loop of automated technologies~\cite{bainbridge1983, Strauch2018, Baxter2012, Fox2023}.

While the issues surrounding automation conceptualization  have been widely studied, less is known about the specific innovation practices that lead to those issues.
Prior HCI works study technology innovation practices within organizations \cite[e.g.,][]{GYLDENKAERNE2024, Fox2024_Humane, George1990, Yildirim2024, Waddell2024}, and together they show that automation implementations are prolonged and complex processes, ``a long and winding road'' (cf.~\cite{GYLDENKAERNE2024}), requiring sociotechnical expertise to configure systems and work practices together \cite{GYLDENKAERNE2024, Fox2024_Humane, Zajac2025}. 
However, these works primarily examine \textit{implementation} processes ---the point at which systems are deployed and integrated into work practices--- leaving underexplored the upstream dynamics through which automation is \textit{selected}, \textit{adapted}, and \textit{tested} before deployment, steps that, according to \citet{DAMANPOUR2006_inn}, are crucial for organizations that adopt externally sourced innovations. 
\textbf{This focus leaves unexamined how decisions made during these earlier phases, such as how innovation practitioners envision automation, which supplier solutions they choose, or what they prioritize in pilot tests, shape the human-automation arrangements that workers later inherit.} Therefore, we propose the following as a first research question:

\begin{quote}    

\textbf{RQ1. }\textit{How do innovation practitioners translate organizational
automation visions into implementable pilot projects, and what role do their conceptualization, framing, and evaluation practices play in the process?}

\end{quote}
\hfill

To address RQ1, this paper analyzes the end-to-end scope of innovation practices taking place in a concrete organization,  Amsterdam Airport Schiphol, a major European airport currently engaged in the long-term \textit{``Autonomous Airside Operations 2050'' }program\footnote{https://www.schiphol.nl/en/innovation/blog/an-autonomous-airport-in-2050/ (Last accessed: 2026-08-01}.
The program is aimed at automating its airside\footnote{The airside is the external, security-restricted area of an airport terminal where aircraft movements and supporting ground operations take place \cite{airside}.}\label{airside} and terminal processes, including, among many other projects, Passenger Boarding Bridge operations or the assistance of passengers through autonomous wheelchairs. These projects are designed, steered, and managed by the organization's innovation department and innovation practitioners.

Furthermore, we aim to identify which specific elements of these early innovation stages could be targeted to support automation adoption, along with directions for designing future interventions. 
Thus, we critically engage with the innovation practices in Amsterdam Airport Schiphol  to identify the aspects that lead to issues in the worker-automation arrangements workers eventually encounter. 
Doing so requires a theoretical lens capable of informing not just how automation is designed, but how design-time decisions relate to use-time interactions. 
Existing frameworks such as Levels of Automation~\cite{parasuraman_model_2000, kaber_effects_2004}, Trust in Automation~\cite{lee_trust_2004}, and Distributed Cognition~\cite{hollan_distributed_2000, rogers_distributed_1994} have sought to characterize appropriate human-automation relations, but they tend to treat automation design as a one-time determination, assuming that how a system is designed directly predetermines how it will be used. This overlooks the iterative work that workers perform to adapt systems to their needs and contextual knowledge.

Co-performance theory \cite{kuijer_co-performance_2018, kuijer_automated_2019} addresses this limitation directly by viewing human-machine arrangements as dynamic, contested, and continually redesigned through use. 
It emphasizes the relationship between design time and use time, where practitioners materialize particular ideas about appropriate worker-automation arrangements at design time, which are then enacted and negotiated in practice, recognizing that appropriatenss cannot be determined a priori. 
We therefore use it as our conceptual foundation to address our second research question:

\begin{quote}    
\textbf{RQ2. }\textit{What recursive relationships exist between innovation practices and use outcomes in automation projects, as viewed from a co-performance lens?}
\end{quote}

Our paper contributes the following to the HCI and design research communities: (1) empirical insights on innovation practices that influence automation conceptualization in a specific organization, (2) an analysis of the recursive relationship between innovation and use practices within the automation projects of the organization, based on co-performance theory, and (3) implications for future research and practice, discussing how future work might rethink and guide automation innovation practices in organizations in ways that enable appropriate worker-automation co-performances.

    We adopted a two-pronged approach. 
    We first conducted in-depth interviews with nine innovation practitioners from Amsterdam Airport Schiphol's innovation department who hold prior experience in the automation projects involving its airside and terminal processes.
    We examined their approaches to automation innovation, roles, workflows, and challenges during project conceptualization.
    In parallel, drawing on co-performance research~\cite{kuijer_co-performance_2018, kuijer_automated_2019, van_beek_making_2023, van_beek_everyday_2025, giaccardi_technology_2020}, we identified and articulated a set of five design principles for application in workplace automation contexts.

Our findings revealed challenges in translating automation visions into concrete projects, the influence of contextual constraints and technology suppliers, and a tendency to prioritize on-site technology testing while deferring worker and process considerations to later stages, ultimately resulting in a lack of contextual integration.
Afterwards, we compared these results with the co-performance principles, identifying agreements and tensions.
This analysis helped us disentangle the recursive relationship between automation innovation practices and use practices in an organization, offering 
nuances and new vocabulary regarding the know-how that practitioners embed in automated equipment across project pipelines, and revealing concrete aspects and directions for intervention.

In summary, 
we account for the organizational intricacies of automation conceptualization, discuss tangible artifacts and innovation elements practitioners directly engage with, and propose starting points for alternative approaches that could reframe current innovation practices.
Through our work, we also exemplify co-performance theory in the organizational domain, a context different from its conception, 
and  reflect on the insights this translation brings into understanding the relation between innovation and use practices.


\section{Related Work}

We position our work at the intersection of HCI research on challenges in designing appropriate worker-automation arrangements (\ref{Challenges autom}) and research on innovation practices in automation adoption within organizations (\ref{Org theories}). 
In addition, we review key concepts related to our theoretical lens, co-performance, and outline our rationale for choosing this lens over other approaches to human–automation relations (\ref{sec:co-performance-theory}). 
Because AI and automation overlap substantially in the concepts and implications covered by research on each, we draw on literature from both fields throughout this review, using the terms AI and automation interchangeably unless a distinction is relevant.


\subsection{Challenges in the Conceptualization of Worker-Automation Arrangements in Organizations}\label{Challenges autom}



  HCI research documents factors that shape, challenge, or postpone how appropriate worker-automation arrangements are conceptualized in organizational contexts. This section examines key challenges. 

First, automation and AI are often presented as urgent, \textit{quick fixes} for complex and entangled organizational problems~\cite{Mlynar2022, Gamkrelidze2024, Maibaum2022}, such as labor shortages~\cite{Morosan2022, Aljuneidi2024} or capacity issues~\cite{GomezBeldarrain2025}, promising, among other benefits, to save time, increase efficiency, or help cope with deteriorating working conditions~\cite{Gamkrelidze2024}. 
This reflects a form of \textit{technological solutionism}~\cite{Morozov2013}, 
or the belief that technology \textit{alone} can resolve socio-organizational deficiencies \cite{Gamkrelidze2024}. 
 As a result, initial efforts and investments in automation projects tend to focus on technology development, while underlying organizational issues are postponed \cite{Gamkrelidze2024}  
and  involved workers are treated as an  afterthought~\cite{GomezBeldarrain2025}. 
Consistent with this perspective, \citet{Mlynar2022}  explain that AI is often portrayed as a heroic character, a \textit{``deus ex machina''} that will address apparently unsolvable problems with enormous effects.
Furthermore, this phenomenon is reinforced by technology suppliers, who promote automation as liberating while masking self-interest: they still require human involvement but gain greater control over production \cite{Baur2023}.


Second, \textit{myths of autonomy}~\cite{Bradshaw2013}, socially widespread misunderstandings about what automation is and should be capable of doing, closely influence how it is 
conceptualized in organizations. 
Empirical work shows that misconceptions about automation are prevalent, ``even in environments that are considered at the forefront of automation innovation'' ~\cite[p. 15]{GomezBeldarrain2025}.
In \textit{``Seven deadly myths of autonomous systems''}, \citet{Bradshaw2013} alert about how such beliefs misguide developers and lead to misplaced foci and goals in the conceptualization of worker-automation arrangements. 
These include, for instance, viewing automation as unidimensional (Myth 1), treating ``full automation'' as an always possible and desirable goal (Myth 7), or using machines as simple substitutes of human capabilities (Myth 6).
Along similar lines, \citet{Riek2025} expose and refute five myths that frequently masquerade as arguments for automation, 
highlighting how they 
constrain technologists and project leads
from critically questioning their own work or serve to justify its purpose~\cite{Gode2020}. 
These dynamics lead to user-hostile worker-automation arrangements and unintended consequences, such as  the well-documented \textit{ironies of automation} \cite{bainbridge1983}, 
 which refer to paradoxical effects of worker-in-the-loop automation setups, including skill degradation, system failures, and misplaced responsibilities ~\cite{Baxter2012, Strauch2018, Shukla2025}. 

Third, automation conceptualizations often rely on representations of workers, work practices, and working conditions~\cite{Breuer2023, Valles-Peris2020}, which can be reductionist~\cite{Akridge2024, Riek2025, Bradshaw2014_2} or mismatched~\cite{Fox2023}.
%
%
\textcolor{black}{
This is caused by the socioeconomic distance between workers and the practitioners designing 
automation~\cite{Akridge2024}, as well as from the often \textit{invisible labor}
performed by workers daily\footnote{Invisible labor refers to essential work activities ---such as informal problem-solving, workarounds, and coordination efforts--- that remain unrecognized or undocumented in formal job descriptions and process specifications~\cite{Star1999, Suchman1995}.}.}
%
Consequently, resulting worker-automation arrangements tend to include overly simplistic imitations of workers' tasks, and fail to address the full range of circumstances they encounter~\cite{Akridge2024}.
Yet, workers remain essential through supervision, maintenance, command, or repair activities~\cite{Delfanti2021, Iantorno2022}, and the need to recognize their continued presence in automated systems has been widely emphasized~\cite{GomezBeldarrain2025, Fox2023, Joo2024}.
For instance, \citet{Fox2023} introduce the notion of \textit{patchwork}, to describe human labor ``that occurs in the space between what AI purports to do and what it actually accomplishes'' (p. 1).
This involves complex acts of integration, troubleshooting, and improvisation to compensate for ill-designed automation~\cite{Fox2023, MontielValle2024, Fox2024_Humane} or to \textit{extend} machinery and its reach~\cite{Delfanti2021}. 
These tasks create new forms of labor that necessitate further study~\cite{Fox2023, Moradi2025} and diverge from the ``upskilling'' promises often associated with automation.



This section illustrates that automation design and deployment in organizations often rely on reductionist misconceptions, leading to poorly conceived, harmful human-automation arrangements. Without alternative conceptualization approaches, these problematic perspectives will likely persist. Therefore, this study analyzes innovation practitioners' current practices and challenges, and explores how a co-performance lens could offer valuable nuance.


\color{black}


\subsection{The Interplay Between Organizations and Automation: the Role of Innovation Practitioners}\label{Org theories}

To study automation innovation in Amsterdam Airport Schiphol, we use~\citet{DAMANPOUR2006_inn}'s framing of \textit{innovation-adopting} organizations.
Innovation-adopting organizations use externally sourced innovations, in contrast to \textit{innovation-generating} organizations. 
They must select, adapt, test, and implement external innovations in their own context \cite{DAMANPOUR2006_inn}, and thus are involved in specific processes and practices that often
rely on innovation teams
~\cite{Johnsson2017_highperf, hellstrom2002, Klitsie2019}.

Related to that, we noticed that only a few HCI works study technology innovation practices within organizations \cite[e.g.,][]{GYLDENKAERNE2024, Fox2024_Humane, George1990, Yildirim2024, Waddell2024}. 
Within the topic of AI and automation innovation, \citet{GYLDENKAERNE2024} identify 14 interdependent sociotechnical tactics for implementing Machine Learning (ML) in a hospital context.
The tactics (e.g., \textit{``decoupling the project from daily clinical work,''} \textit{``providing proof of concept,''} etc.)  span three analytic levels---organization, project, and practice---and blur development and implementation boundaries. 
\citet{Zajac2023} explore clinical implementations of ML in multiple studies~\cite{Zajkac2024, Zajac2025}, while \citet{Fox2024_Humane} examine tensions from the rapid introduction of AI in essential sectors.
Together, these studies show that AI implementation is a prolonged and complex process ---``a long and winding road''~\cite{GYLDENKAERNE2024}--- requiring sociotechnical expertise to configure systems and practices together~\cite{GYLDENKAERNE2024, Fox2024_Humane, Zajac2025}, which we believe extends to automation implementation. 
They offer key findings: system' usefulness depends on configurability~\cite{Zajac2025}, solutions cannot be finalized outside deployment contexts~\cite{Fox2024_Humane, GYLDENKAERNE2024, Zajac2025}, stakeholder misalignment multiplies work~\cite{GYLDENKAERNE2024}, and workers must be treated as domain experts~\cite{Fox2024_Humane, Zajac2025}---insights aligning with our theoretical lens of co-performance (Section~\ref{sec:co-performance-theory}).

However, these works share a common scope: they primarily examine the \textit{implementation} process of AI and automation technologies---the point at which systems are deployed and integrated into work practices. This focus leaves underexplored the upstream dynamics through which automation is \textit{selected, adapted, and tested} before deployment (cf.~\cite{DAMANPOUR2006_inn}).  
Critically, decisions made during these earlier phases---how innovation practitioners conceptualize automation, which supplier solutions they choose, what they prioritize in pilot tests---shape the human-machine arrangements that workers later inherit. In this work, we address this gap by analyzing the full scope of innovation practitioners' work, from vision translation through piloting. 

 This focus on innovation practitioners draws on the social construction of technology (SCOT) framework~\cite{Pinch1984}, which describes technology development as multi-directional and influenced by relevant social groups. 
Following the categorization of \citet{Humphreys2005}, we label innovation practitioners as \textit{producers}, since they shape technology through design and investment---a group that remains understudied in automation contexts~\cite{GomezBeldarrain2025} compared to \textit{users} (workers)~\cite{Ma2023, Spektor2023DIS}, \textit{advocates} (policy makers)~\cite{Eskenazi2023, Mladenovic2020}, and \textit{bystanders} (cultural critics, general public)~\cite{Bassett2019}, despite their pivotal role.
Relatedly, our epistemological grounding lies in \textit{sociomaterial} theories~\cite{Bailey2022, Leonardi2011, Leonardi2013, Orlikowski1992, Orlikowski2008, Orlikowski2007}, 
which posit that material and social aspects within organizations are \textit{constitutively entangled}---
rejecting approaches that treat them as separate~\cite{Jackson2002, Leonardi2013, Leonardi2008}. 
For our analysis, this means focusing on the co-constitution of material artifacts (i.e., automation technology) and human actions (i.e., innovation practices), where automation both enables and constrains organizational arrangements, while innovation practitioners' tasks and contextual constraints shape how technology is conceptualized and scaled~\cite{Bailey2022}.

\color{black}

\subsection{Co-Performance as a Theoretical Framework for Understanding Automation Appropriateness}\label{sec:co-performance-theory} 

\color{black}


To critically engage with innovation practices in our context, Amsterdam Airport Schiphol, and identify aspects that might lead to issues in downstream worker-automation arrangements, we ground our analysis in a theoretical framework that helps us understand human-automation appropriateness.
This section outlines our rationale for choosing the co-performance lens and summarizes its key dimensions.



To ground our selection, we focus on frameworks that offer theoretical lenses for understanding human-automation relations, rather than design guidelines (e.g., Human-AI Interaction~\cite{amershi_guidelines_2019}), which provide practical heuristics without explanatory theory, or domain-specific applications (e.g., Human-Robot Teaming~\cite{gombolay_computational_2017}), where concepts like ``teaming'' function as design goals rather than theoretical lenses.
Influential analytical frameworks include Distributed Cognition, which analyzes how cognition extends across people, artifacts, and time~\cite{hollan_distributed_2000, rogers_distributed_1994}; Levels of Automation, which systematically models function allocation between humans and machines~\cite{parasuraman_model_2000, kaber_effects_2004}; and Trust in Automation, which examines how trust forms and mediates decisions about automation reliance~\cite{lee_trust_2004}.

Co-performance, drawn from practice-theoretical approaches to HCI~\cite{kuutti_turn_2014}, shares common ground with each: like distributed cognition, it examines joint human-machine accomplishment; like levels of automation, it addresses task distribution; and like trust research, it attends to relationships that develop over time. Table~\ref{tab:theoretical-comparison} summarizes how co-performance relates to these established frameworks. 
However, co-performance differs in a crucial respect.
Rather than treating human–automation task arrangements as predetermined at design time, it understands them as negotiated through situated practice and therefore views the relationship between design and use as \textit{recursive}. Human–machine arrangements are consequently understood as dynamic, contested, and continually reshaped through use. While the established frameworks each provide valuable conceptual insights that co-performance does not fully encompass, our research focuses specifically on how design-time decisions and use-time outcomes recursively shape one another. Co-performance addresses this dynamic explicitly, making it particularly well suited for studying innovation practices across their full temporal and organizational scope.


\begin{table*}[t]
\caption{Comparison of theoretical lenses for human-automation relations. Co-performance shares common ground with each established framework but treats appropriateness as negotiated through situated practice, making it suited for critiquing the relationship between innovation practices and use practices in automation projects.}
\label{tab:theoretical-comparison}
\small
\renewcommand{\arraystretch}{1.3}
\begin{tabularx}{\textwidth}{@{}>{\raggedright\arraybackslash}p{2.8cm} >{\raggedright\arraybackslash}X >{\raggedright\arraybackslash}X@{}}
\toprule
\textbf{Framework} & \textbf{What it offers} & \textbf{How co-performance differs} \\
\midrule

Distributed Cognition~\cite{hollan_distributed_2000, rogers_distributed_1994} & 
Goes beyond the individual to examine how humans and machines jointly accomplish tasks; analyzes information processing, representations, and flows through systems & 
Focuses on embodied learning in everyday practice and the ongoing negotiation of appropriateness during performance---a distinguishing feature treated less explicitly in distributed cognition \\
\addlinespace[0.8em]

Levels of Automation~\cite{parasuraman_model_2000, kaber_effects_2004} & 
Studies how tasks are distributed across humans and machines; early work treated tasks as static functions assigned at design-time; later adaptive automation work addresses dynamic, context-sensitive allocation & 
Goes further: views appropriate assignment of tasks as negotiated and (re)designed at use-time, not merely adapted based on predefined criteria \\
\addlinespace[0.8em]

Trust in Automation~\cite{lee_trust_2004} & 
Studies development and change of relationships between humans and machines over time; focuses on individual psychological attitudes and calibration of trust to match system capabilities & 
Adopts a practice-theoretical lens focusing on how humans and machines \textit{jointly} negotiate what is appropriate in specific situations, rather than calibrating individual attitudes toward fixed capabilities \\

\bottomrule
\end{tabularx}
\end{table*}

The notion of co-performance is introduced by \citet{kuijer_automated_2019} and \citet{ kuijer_co-performance_2018} in the context of smart home appliances, domestic practices, and human-computer interaction,  which we summarize below. 
This body of work has since been widely used to frame interactions between humans and intelligent systems across various domains, including not only domestic \cite{Kuijer2022} and smart home settings \cite{van_beek_everyday_2025, van_beek_making_2023, Viaene2021}, but also algorithmic systems and digital twins \cite{Turtle2025}, everyday IT services \cite{KimLim2019}, explainable AI \cite{Nicenboim2022}, or the entanglement between professional artists and LLMs \cite{Bomba2024}.


Co-performance is based on practice theory \cite{Schatzki2002, Shove2012, Reckwitz2002}, 
and shifts focus from the supposed autonomy of artifacts to seeing automated artifacts as capable of learning and performing next to people \cite{kuijer_co-performance_2018}. 
This view extends traditional practice-theoretical approaches \cite{Schatzki2002, Shove2012, Reckwitz2002} in their understanding of `the material': rather than treating all technologies (e.g., a chair, a spoon, or an autonomous wheelchair) as similarly inert, subordinate, and interchangeable, it considers the performance of `artificial doings' alongside human performers~\cite{kuijer_automated_2019, kuijer_co-performance_2018}.
Agency becomes more than the mere capacity for causal action (e.g., transporting a person), but also the contribution to the performance of practices across time and space (e.g., navigating and accompanying passengers with reduced mobility to their flight boarding area), with automated artifacts as integrators of materials, competences, and meanings~\cite[p.~3]{kuijer_co-performance_2018}. 
    Importantly, 
    human and artificial agency are not separate categories but emerge through co-performance itself~\cite{kuijer_co-performance_2018}. Automation represents one set of techniques through which artificial agency manifests and is the focus of this paper.

Co-performance highlights that differences between the capabilities of humans and automated artifacts matter and determine the division of work in any given practice~\cite[pp. 200, 202--203]{kuijer_automated_2019}.
Automated artifacts embody practical knowledge and know-how, distinct for their ability to incorporate accumulated learning across generations~\cite[pp. 205--206, 209]{kuijer_automated_2019}.
Thus, when tasks are transferred from humans to automated artifacts, they are transformed due to this same difference in capabilities, leading to both losses \textit{and} gains in the overall practice. 
That is to say, machines are rarely simply `better' at  tasks previously performed by humans; rather, they perform them \textit{differently}, and learn differently too, accumulating knowledge across generations at speeds humans cannot match, yet lacking the situated enculturation through which humans develop contextual judgment~\cite[pp. 205--206, 209]{kuijer_automated_2019}.

Moreover, 
the appropriateness of any given co-performance `regime' cannot be determined a priori, it emerges through the dynamic and contested reality of everyday life~\cite{van_beek_everyday_2025, kuijer_co-performance_2018}.
There is a recursive relationship between design and use; design continues at `use time.' While design decisions are based on long-term historic and culturally varied processes in everyday life, these decisions do shape but cannot fully determine practices at use time~\cite[p. 9]{kuijer_co-performance_2018}. This nuance of co-performance regarding the dynamic relation between the design and use of automated artifacts is key in our research.
Related to that, as \citet{van_beek_everyday_2025} highlight, while breakdowns, frictions, and malfunctions are typically treated as undesirable in human-centered design (except where friction is used tactically~\cite[e.g.,][]{Gould2021, Haliburton2024}),
co-performance considers crises as moments of opportunity for the reconfiguration of task arrangements~\cite{van_beek_everyday_2025, van_beek_making_2023}.
Finally, a co-performance perspective reconceptualizes the notion of interface not as a static `control panel' but as a dynamic matching of peoples and things enacted \textit{through} joint doings.
Consider, for instance, an airport worker accompanying an autonomous wheelchair through a crowded terminal (cf.\ Section~\ref{sec: background}): the interface is not merely the device's control panel, but emerges through their joint navigation. 

Taken together, co-performance offers a distinctive theoretical lens for studying human-automation arrangements and innovation practice; unlike frameworks that fix function allocation at design-time, co-performance foregrounds the recursive relationship between design and use, the role of situatedness in everyday practice, and the productive potential of breakdowns in reconfiguring arrangements. 
  In this paper, we draw on this foundational literature~\cite{kuijer_automated_2019, kuijer_co-performance_2018, van_beek_everyday_2025, van_beek_making_2023, giaccardi_technology_2020} to 
  articulate a set of design principles, which we then use as a conceptual lens to analyze the innovation practices in our automation context (cf. Table~\ref{tab:co-performance-principles}).
\textcolor{black}{In doing so, we shift the scale and nature of ``practice'' from domestic social practices (e.g., heating, laundering) to organizational work practices (e.g., baggage handling, aircraft servicing), following prior attempts~\cite{van_beek_everyday_2025} to move co-performance from domestic to organizational settings (cf. Section~\ref{disc: co-perf}).} 


\color{black}

\section{Background: Automation Innovation in an International Airport} \label{sec: background}





\begin{figure*}[htb]
    \centering

    \includegraphics[width=1.0\textwidth]{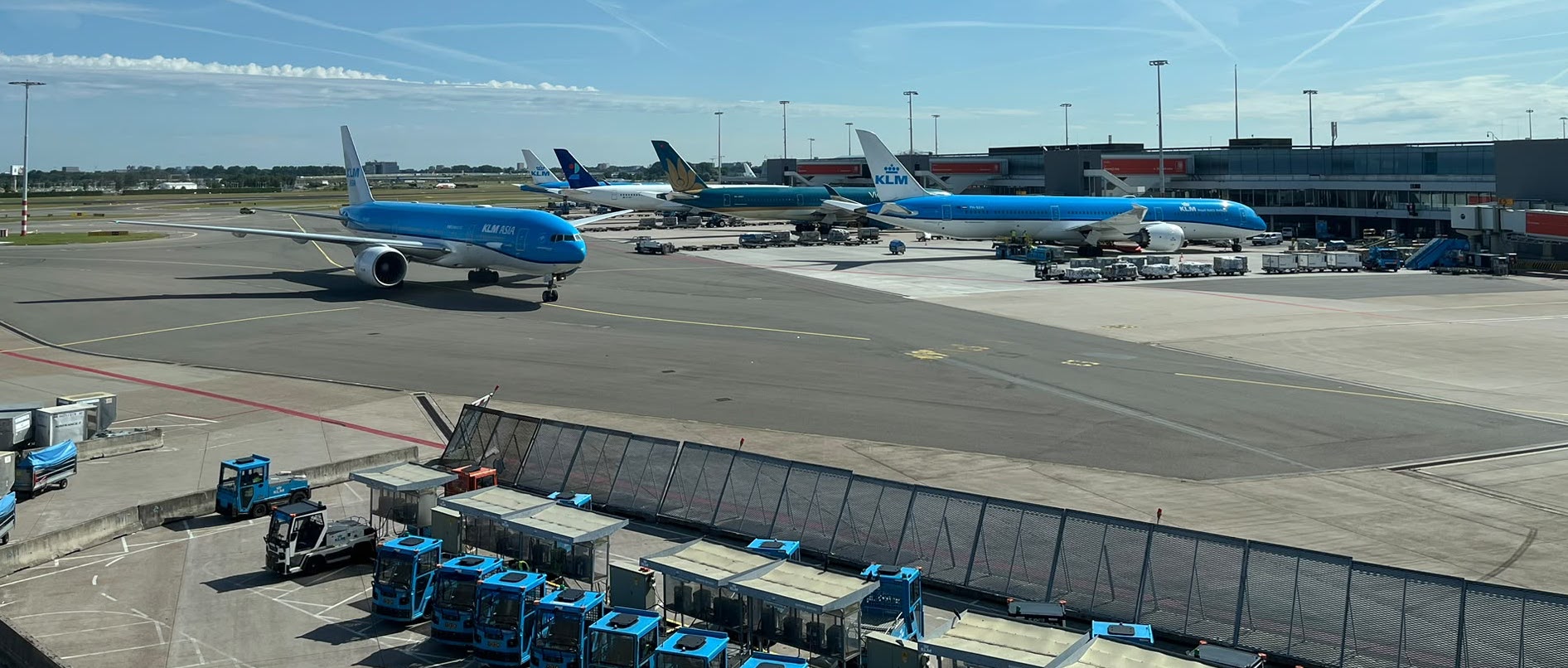}

    \caption{View of Amsterdam Airport Schiphol's airside, with the terminal at the background. Automation projects are intended to take place both in the airside and terminal areas.
    \textit{Image courtesy of  Silvia Rey Abeijón, The Royal Schiphol Group.}}

    \Description{Representative airside scene illustrating the operational environment considered in this work. The image captures commercial aircraft at gates, a taxiing aircraft, ground support infrastructure, and apron markings, providing the context in which the proposed automation solutions are intended to be deployed.}
    \label{fig:Fig_Airside}
\end{figure*}

\begin{figure*}[htb]
    \centering

    \begin{minipage}[t]{0.49\textwidth}
        \centering
        \includegraphics[width=\textwidth]{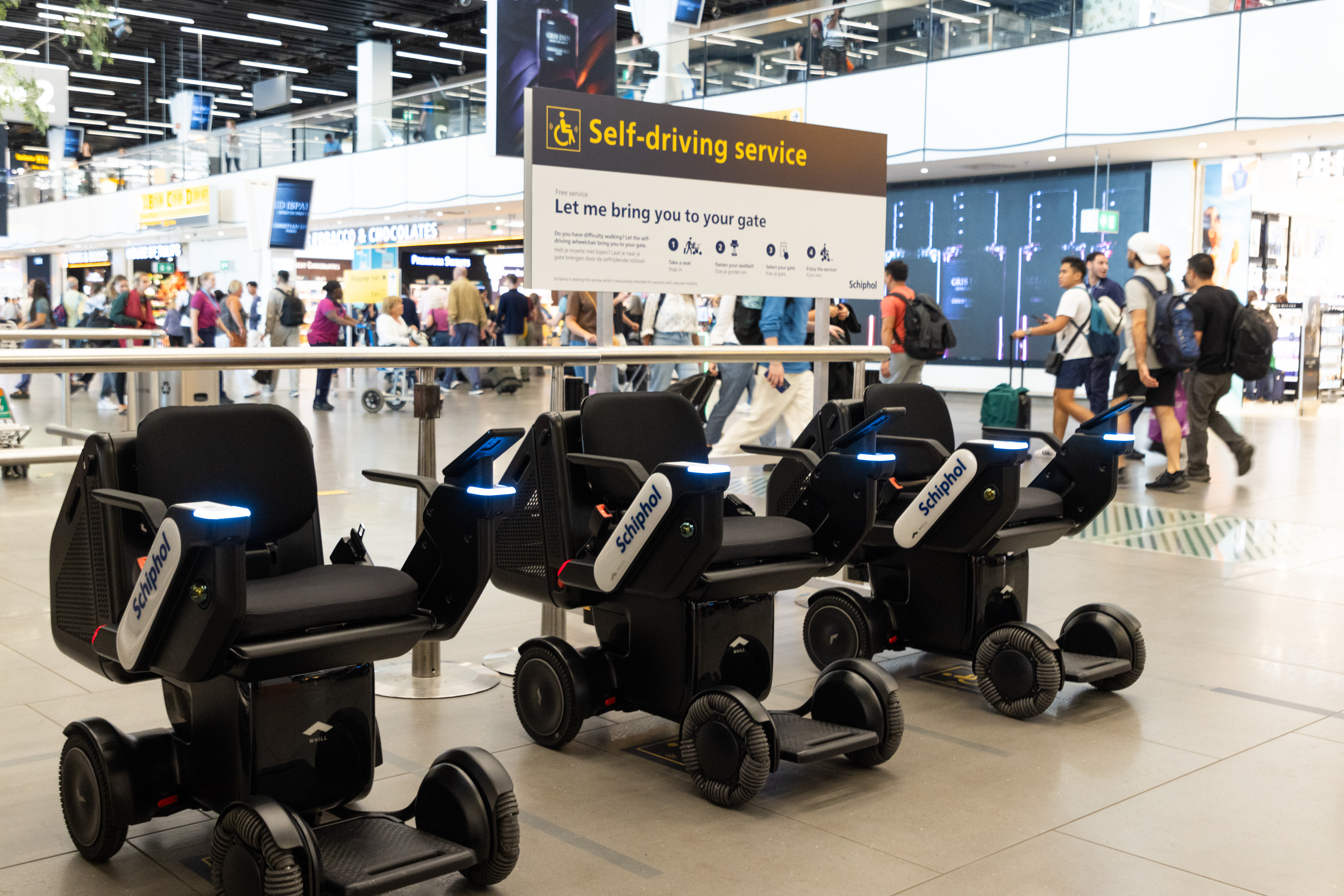}
        \label{fig:b_Bus}
    \end{minipage}
    \hfill
    \begin{minipage}[t]{0.49\textwidth}
        \centering
        \includegraphics[width=\textwidth]{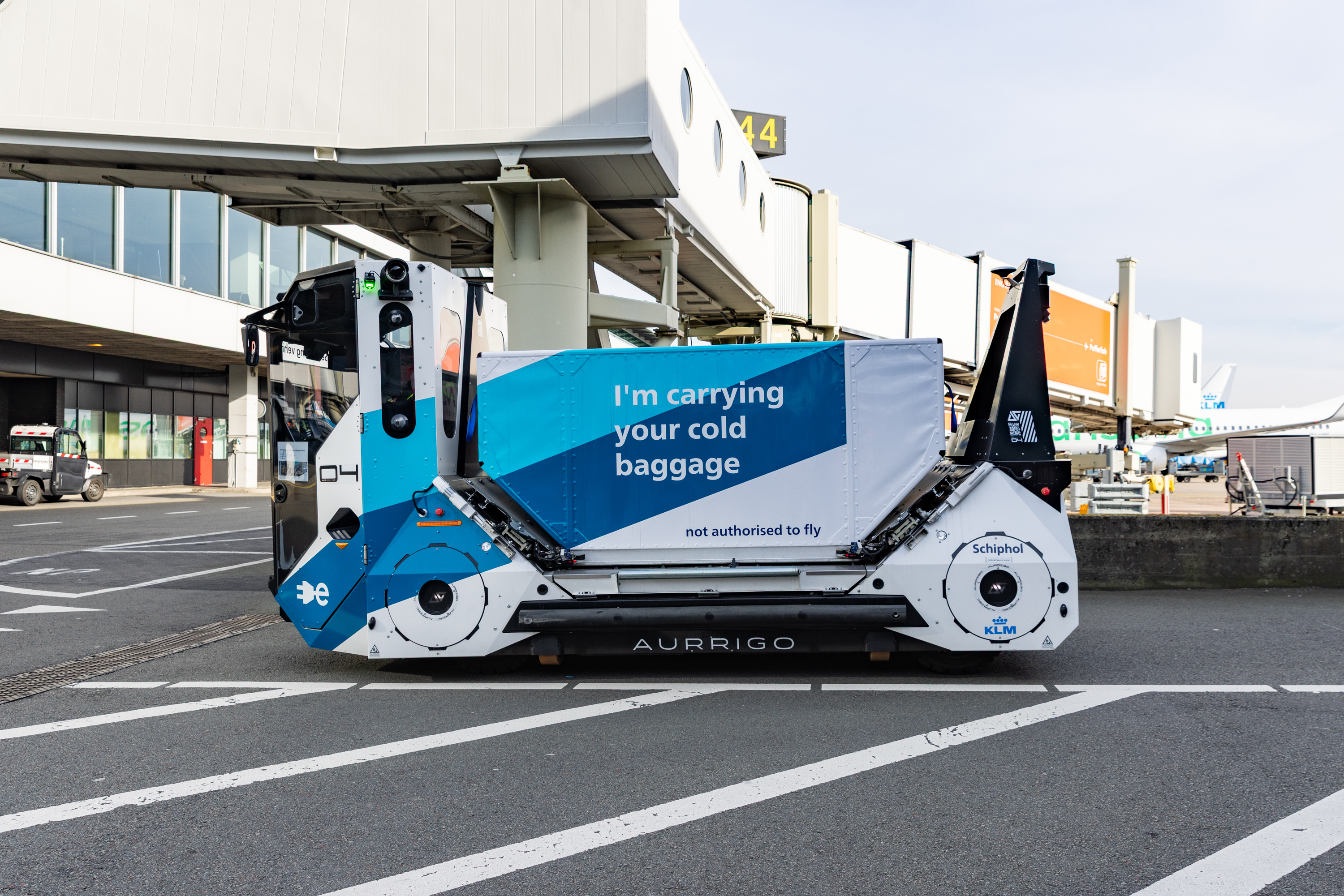}
        \label{fig:c_cobot}
    \end{minipage}
\vspace{-1em}
    \caption{Images taken during on-site pilot tests, illustrating two automation projects conducted by the innovation department of Amsterdam Airport Schiphol: (left) the autonomous wheelchair project, aimed at transporting passengers with reduced mobility in the terminal, and (right) the autonomous baggage handling vehicle project  aimed at transporting baggage in the airside from incoming aircraft to buffer areas. \textit{ Images courtesy of The Royal Schiphol Group.}}

    \Description{Two images visualizing automation projects. The image on the left shows three autonomous wheelchairs parked inside the terminal. The image on the right shows an autonomous vehicle that is pulling baggage containers from an aircraft to a buffer area.}
    \label{fig:Fig2}
\end{figure*}

We study automation innovation in a specific context, Amsterdam Airport Schiphol, a major international airport in Western Europe. 
Since 2020, the organization has undertaken a 30-year automation program to address workforce scarcity, airport capacity, and sustainability challenges.
The program envisions a fully autonomous airside (i.e., the external airport area dedicated to aircraft movements, Figure \ref{fig:Fig_Airside}) by 2050, along with selected terminal processes, all of which are organized within a roadmap.
Examples of those autonomous operations include baggage sorting robots, autonomous vehicles for passenger and baggage transportation, snow removal services, and autonomous wheelchairs for passengers with reduced mobility. See Table~\ref{tab:projects overview} in the Appendix for an overview of conducted projects, Figure~\ref{fig:Fig2} for illustrative examples.

The airport’s innovation department is responsible for executing the strategy and managing the initial stages of these automation projects, with a dedicated team of around 10 innovation practitioners. 
Their main tasks include: 
(1) identifying innovation opportunities, (2) scouting for suppliers and technology developers, (3) soliciting tenders,
(4) conducting pilot tests, and (5) proposing detailed automation concepts for scaling.
They must also align these projects across departments, engaging diverse stakeholders (e.g., service owners, management, air traffic authorities).
So far, conducted projects have involved long testing timelines, implementation difficulties, and efforts to align suppliers and airport stakeholders; many projects were discontinued after lengthy pilot tests. 
In this paper, we focus on innovation practices, to examine the recursive relationship between innovation and use practices,  highlighting how these dynamics may result in particular worker–automation arrangements.
The first author of this paper is an embedded researcher in the Innovation Department [see Section \ref{positionality}].

\color{black}

\section{Method}\label{sec:method}


\subsection{Overall Approach}\label{sec:overall-approach}

This research aims to study automation innovation practices in an organizational setting. Our qualitative-interpretive approach recognizes that understanding organizational practices is mediated through social and subjective filters, viewing knowledge as contextual yet meaningful within specific settings. This enables us to capture the complexity and nuance of how innovation practitioners conceptualize automation within their organizational context, going beyond surface description, in line with recent HCI  
works employing similar approaches \cite{alfrink_contestable_2023-2, claisse_keeping_2023, offerman_rediscovering_2025}. 
\textcolor{black}{
We gather empirical insights through semi-structured interviews with nine innovation practitioners from a major European airport undergoing automation transformation, and use co-performance principles from prior work as a conceptual lens to analyze these insights.
Our single-site focus enables deep contextual understanding of how organizational intricacies shape automation conceptualization (detailed rationale in Section~\ref{sec:interview-study}).
}

\begin{figure*}[htb]
    \centering
        \includegraphics[width=1.0\textwidth]{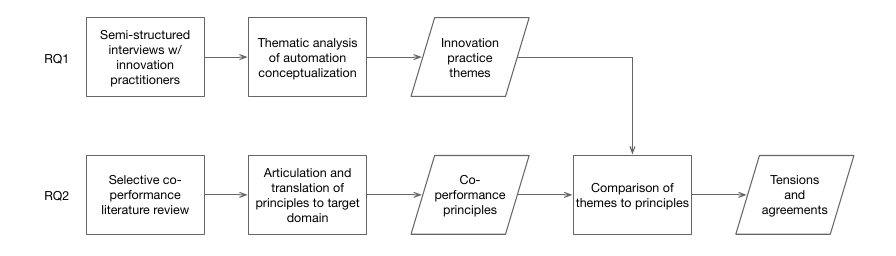}
        \caption{Overview of our research approach. Two parallel strands address each research question: RQ1 draws on semi-structured interviews analyzed thematically to produce innovation practice themes, while RQ2 synthesizes co-performance principles from selective literature review. These strands converge in a comparative analysis that identifies tensions and agreements between current practices and co-performance theory. Rectangles represent research activities; parallelograms represent inputs and outputs.
        }
        \Description{A flowchart depicting the paper's two-strand methodology. The top strand, labeled RQ1, shows three connected elements flowing left to right: 'Semi-structured interviews w/ innovation practitioners' (parallelogram) leading to 'Thematic analysis of automation conceptualization' (rectangle) leading to 'Innovation practice themes' (parallelogram). The bottom strand, labeled RQ2, shows: 'Selective co-performance literature review' (parallelogram) leading to 'Articulation and translation of principles to target domain' (rectangle) leading to 'Co-performance principles' (parallelogram). Both strands converge at a central rectangle labeled 'Comparison of themes to principles,' which outputs to a final parallelogram labeled 'Tensions and agreements.}
        \label{fig:Researchdiagram}
\end{figure*}

We employed reflexive thematic analysis~\cite{cooper_thematic_2012, BraunC} to identify themes from the interviews, compared these \textcolor{black}{against co-performance principles} to identify gaps, and synthesized our reflections in a discussion (Section \ref{sec:discussion}). Member checking with three participants validated our interpretations. The study received institutional ethics approval. Figure~\ref{fig:Researchdiagram} illustrates the complete process.

\subsection{Interview Study (RQ1)}\label{sec:interview-study}

\begin{table*}[htb]
  \centering
  \caption{Overview of the interview participants. We describe their roles in automation projects and their professional experience in years. To avoid re-identifications, we do not specify participant IDs.}
  \label{tab:participants}
  \begin{tabularx}{\textwidth}{lrcX}
    \toprule
    \textbf{Role} & \textbf{Years of experience} & \textbf{Number} & \textbf{Description} \\
    \midrule
    Innovation leads & 11--20 & 3 & Manage visions and project portfolios, and lead teams of innovation~practitioners. \\
    \addlinespace
    Innovation practitioners & 1--10 & 5 & \multirow{2}{\hsize}{In charge of specific projects assigned by innovation~leads.} \\
    & 11--20 & 1 & \\
    \bottomrule
  \end{tabularx}
\end{table*}


\subsubsection{Recruitment}
Participants were recruited through purposive sampling within Amsterdam Airport Schiphol's innovation department, targeting innovation practitioners actively involved in automation projects. 
Since the first author is embedded in the context (see Section \ref{positionality}), 
initial invitations were sent directly, followed by snowball sampling to identify additional participants with relevant experience in automation conceptualization and implementation. 
We aimed to interview about 8--10 innovation practitioners to understand their approach and workflows in automation projects and identify challenges in automation conceptualization. 

\subsubsection{Participants}
We conducted interviews with 9 participants
from the airport's innovation department; see the overview in Table ~\ref{tab:participants}. 
The sample included innovation practitioners in charge of concrete projects (n=6) and leads managing visions and automation portfolios (n=3)\footnote{So far we referred to them collectively as \textit{innovation practitioners} since they are all part of the innovation team of Amsterdam Airport Schiphol, work on automation innovation, and mainly differ in their responsibilities and seniority\label{footnote}. }. 
Participants' professional experience ranged from 1--20 years, with 5 participants having 1--10 years of experience and 4 participants with 11--20 years. 
All participants were actively involved in projects within the airport's Autonomous Airside Operations program at the time of the study and agreed to participate through signed consent forms. These projects include implementations of autonomous baggage vehicles, autonomous buses, baggage-handling robots, autonomous passenger boarding bridges, and autonomous ground power units, among others (see Table \ref{tab:projects overview});
to preserve participant anonymity, we do not link individual participants to their specific projects.

\subsubsection{Procedure}
Semi-structured interviews lasted 45--75 minutes. Following participants' preference, they were mostly conducted via video conferencing, with the exception of two in-person interviews.
With the aim of understanding the aspects and dynamics that influence automation conceptualization during innovation projects, our interview guide went from more general topics (e.g., their roles, projects, and workflows) to more concrete aspects of innovation practitioners' projects (e.g., fallback design, tools)  (see Appendix \ref{Appendix:interview_guide}): 
first, we asked participants about their roles and the automation projects they participated in.
Second, we asked them to describe 
their processes for translating organizational visions into those concrete automation initiatives.
Third, we asked them to explain how proof of concepts are decided upon, scheduled, and conducted, including their
approaches to conceptualizing human-automation collaboration and fallback situations. 
Lastly, we asked them about current resources they use to support their practices, as well as their wishes for new tools or support. 
Participants were asked to describe their workflow, decision-making processes and team dynamics, and show examples of resources they use during project development. Follow-up questions helped us uncover details about these elements.
All interviews were audio-recorded with consent and transcribed verbatim for analysis. 

\begin{table*}[!h]
\centering
\caption{
    \textcolor{black}{
    Co-performance design principles synthesized from foundational literature~\cite{kuijer_automated_2019, kuijer_co-performance_2018, van_beek_everyday_2025, van_beek_making_2023}. These principles operationalize co-performance as a lens for designing with artificial agency (Section~\ref{sec:co-performance-theory}), reframing artifacts not as replacements for human agency, but as co-performers in co-performative relationships, learning to enact practices together in contextually appropriate ways. In this paper, we apply these principles specifically to automation contexts. Together, they offer HCI practitioners a theoretical foundation for designing systems that remain responsive to the situated, contested, and evolving nature of everyday life.
    }
    }
\label{tab:co-performance-principles}
\begin{tabularx}{\textwidth}{c l X}
\toprule
\textbf{ID} & \textbf{Principle} & \textbf{Description} \\
\midrule
A & Design for recursive adaptation & 
    Acknowledge that design is recursive---continuing through use practice and professional design practice~\cite{kuijer_co-performance_2018}. 
    Design openness into systems to enable dynamic rearrangement of human-technology relations~\cite{van_beek_everyday_2025}.
    Support `everyday designers' who modify and reconfigure systems during use, recognizing their role as co-designers~\cite{kuijer_co-performance_2018}. \\
\addlinespace
B & Delegate according to capabilities & 
    Distinguish between polymorphic actions (requiring social awareness and contextual judgment) and mimeomorphic actions (rule-based, consistent operations) when allocating tasks~\cite{kuijer_automated_2019}.
    \textcolor{black}{Recognize that capabilities evolve differently: machines accumulate learning faster across generations and excel at consistent operation, while humans develop contextual judgment through enculturation~\cite{kuijer_automated_2019, kuijer_co-performance_2018}.
    Avoid delegating socially complex judgments to artificial performers that lack this situated experience~\cite{kuijer_automated_2019}.} \\
\addlinespace
C & Enable mutual responsiveness & 
    Design for responsiveness in both human and artificial performers rather than optimizing fixed co-performance arrangements~\cite{van_beek_everyday_2025, giaccardi_technology_2020}. 
    Enable multiple forms of reconfiguration: manual interventions, {socio-}material settings adjustments, and systems reprogramming~\cite{van_beek_everyday_2025}.
    Recognize that appropriateness emerges through situated practice rather than predetermined design decisions~\cite{van_beek_everyday_2025}. \\
\addlinespace
D & Design productive friction & 
    Reframe everyday crises as opportunities for improved co-performances rather than design failures to eliminate~\cite{van_beek_everyday_2025, van_beek_making_2023}.
    Design minor, productive frictions that enable learning without causing excessive distress~\cite{van_beek_making_2023}.
    Create mechanisms for resolving conflicts about appropriate practice between human and artificial performers~\cite{van_beek_everyday_2025}. \\
\addlinespace
E & Support co-learning over time & 
    Design for the extended time required to enact appropriate {co-performances} between humans and technologies~\cite{van_beek_making_2023}.
    \textcolor{black}{Account for different temporal rhythms: human learning unfolds through experience, while machine learning accumulates across generations~\cite{van_beek_making_2023}.
    Support mutual adaptation over time as human and artificial performers learn to co-perform together~\cite{van_beek_making_2023}.} \\ 
\bottomrule
\end{tabularx}
\end{table*}

\subsubsection{Analysis}
The data we worked with are the interview transcripts (n=9). 
We based our overall approach on reflexive thematic analysis~\cite{braun_can_2021, braun_conceptual_2022, braun_reflecting_2019, braun_using_2006, cooper_thematic_2012},  with  researchers having an active role in the analysis. 
We first transcribed audio-recordings,
then cleaned and anonymized the transcripts, and familiarized ourselves with the data by going through the interviews once more.
We employed an iterative coding process.  
The first author started this process by performing open coding on 30\% of the transcripts (n=3), generating an initial set of codes that stayed close to participants' explicit accounts, capturing data patterns inductively related to innovation practices and their relation to, effect on, or consideration of later worker-automation arrangements.
The first author grouped the codes into similar categories, and the author team engaged in joint reflection, through discussion meetings, on what the meanings of the categories were at a latent level, resulting in a candidate set of themes.
Consistent with the interpretive, reflexive nature of this method, disagreements were resolved through discussion and used to sharpen and refine the developing themes.
The first author then coded the remaining transcripts, using the preliminary themes as a guide, but allowing for revision of themes as coding progressed.
Finally, the author team discussed the final set of themes against the full dataset, refining their definitions and names, before finalizing the themes prior to write-up.
A description of the authors' positionality in relation to this research is provided in Section \ref{positionality}.

\subsubsection{Credibility Strategies} \label{credibility strategies}

As already mentioned, we had frequent 
discussions (9 sessions of one hour each) within the author team to ensure a thorough analysis. 
We used reflexivity to account for our particular positions and how these might affect our analysis.
We used member checking with 3 participants---sharing a draft report of our findings with participants for feedback---to ensure our analysis reflects our participants' views  (Section~\ref{sec:overall-approach}). 

\subsubsection{Statement of Positionality} \label{positionality}

The authors are HCI and design researchers with backgrounds in Human-AI Collaboration, Human-Centered AI, Complex Environments, and Design Engineering. This interdisciplinary perspective shapes our approach to studying human-centered technology implementation in complex organizational settings.
We approach this research from a position favoring responsible automation adoption, that addresses diverse stakeholder needs through iterative, participatory processes. This position may have made us more likely to notice and positively interpret practices aligned with participatory, worker-centered automation.
To mitigate researcher bias, rather than relying solely on our own normative judgments, we apply co-performance theory as a lens to examine automation appropriateness.

The first author maintains a non-contractual, embedded research position at Amsterdam Airport Schiphol, spending one to two days per week within the innovation department responsible for automation initiatives and participating in regular meetings and team activities. This position provides direct observation of departmental routines, ongoing projects, organizational culture, and privileged access to innovation practitioners and the broader automation transformation context. 
At the same time, this embeddedness and ongoing relationship with the team risk socially desirable responses from participants, and may make it harder for the first author to draw critical interpretations of the team's practices. These potential biases were actively addressed through the credibility strategies described in Section \ref{credibility strategies}.

We acknowledge that interviewing innovation practitioners captures only one side of this complexity, and that additional stakeholders, including workers, should be considered in subsequent, complementary research.

\subsection{Translating Co-Performance Principles to Workplace Automation (RQ2)}
\label{sec:co-performance-method}

We drew on co-performance literature to articulate a set of principles to be used as a conceptual lens for analyzing workplace automation practices (Table~\ref{tab:co-performance-principles}). Our aim was not to derive new principles, but to translate and operationalize existing ones for this organizational context.

First, we conducted a selective review of foundational co-per\-formance work (cf. Section~\ref{sec:co-performance-theory}), including \citet{giaccardi_technology_2020, kuijer_automated_2019, kuijer_co-performance_2018, van_beek_everyday_2025, van_beek_making_2023}. 
Second, we examined how the principles articulated in this literature may travel into the domain of workplace automation, identifying those most relevant to organizational innovation. 
Third, the author team iteratively refined the articulation of these principles to ensure clarity, scope, and applicability, undergoing three major iterations with smaller adjustments in between. 
Finally, the second author consolidated team feedback into the set presented here.

This articulation provided the foundation for using co-performance as a conceptual lens for examining the practitioner interview data. Rather than a direct comparison, the analysis sought to surface alignments, tensions, and opportunities between existing co-performance principles and current innovation practices in workplace automation.

\color{black}

To conduct this analysis, the author team systematically examined each co-performance principle against each interview theme, asking: in what ways does this principle align with or diverge from current innovation practices as reflected in this theme?
This analysis operated at multiple levels of granularity. 
\textcolor{black}{For instance, the notion of \textit{openness}---a key aspect of designing for recursive adaptation in co-performance theory---was examined in relation to innovation practitioners' descriptions of vision rigidity. 
In other cases, principles and themes were related at a more holistic level. 
For example, full automation narratives from our interviews stood in direct tension with co-performance's emphasis on delegating tasks according to the distinct capabilities of humans and machines.} 
We classified relationships as either \textit{agreement} (when practices supported or enacted the principle) or \textit{tension} (when practices conflicted with or neglected the principle), with the degree of explicitness determining whether these were marked as \textit{strong} or \textit{partial}. 
Explicit, direct connections were classified as strong agreement or tension, while more subtle or implicit relations were classified as partial. 
Cases where no meaningful relationship could be identified were marked as unrelated. 
The resulting matrix (Table~\ref{tab:co-performance-automation}) synthesizes these assessments---using principle and theme identifiers for compact reference---and structures the analysis presented in Section~\ref{sec: RQ2}.

\section{Findings (RQ1): Innovation Practices that Shape Automation Conceptualization in Amsterdam Airport Schiphol}\label{sec:results} 


The resulting themes reveal how practitioners' approaches to innovation shape automation conceptualization in Amsterdam Airport Schiphol (RQ1). We describe their innovation practices within the long-term \textit{``Autonomous Airside Operations''} program the organization is embedded in. 
\textcolor{black}{The themes in this section are summarized in Table~\ref{tab:rq1-summary}.} 
These results form the foundation for our comparative analysis presented in Section~\ref{sec: RQ2}.

\begin{table*}[htbp]
\centering
\caption{Summary of innovation practice themes shaping automation conceptualization (RQ1). Each theme captures a distinct aspect of how innovation practitioners at Amsterdam Airport Schiphol approach automation innovation, from vision translation through piloting and human considerations.}
\label{tab:rq1-summary}
\small
\begin{tabularx}{\textwidth}{@{}l >{\raggedright\arraybackslash}p{0.25\textwidth} >{\raggedright\arraybackslash}X >{\raggedright\arraybackslash}p{0.25\textwidth}@{}}
\toprule
\textbf{§} & \textbf{Theme} & \textbf{Key findings} & \textbf{Illustrative practitioner~quote} \\
\midrule
\ref{sec:theme-visions} & Vision as a starting~point 
& Visions serve as a ``North Star'' providing direction toward full automation by 2050; however, innovation practitioners perceive them as static rather than iterative. Translation from abstract visions to concrete projects is challenging, and practitioners report feeling ``loneliness'' between leadership and operations.
& \textit{``It's somewhere in the North, but where exactly we will end up? We don't know.''}~(P5) \\
\addlinespace
\ref{sec:theme-projects} & Project development in~practice 
& Projects originate not only from visions but also from opportunistic discoveries and cross-departmental requests. Innovation practitioners face scope creep and must ``protect the scope'' while managing diverse stakeholders. The reality is messy and entangled rather than linear.
& \textit{``People management, involving the right people, at the right time, asking them the right questions [...] is the majority of the work and really the hard part.''}~(P6) \\
\addlinespace
\ref{sec:theme-constraints} & Contextual constraints 
& Solutions are shaped by business demands and defined as direct translations of manual operations (``one-on-one replacement''). Contextual constraints---infrastructure, regulations, budget, operational rules---strongly dictate automation concepts. Innovation practitioners aim to ``mimic'' existing context behavior.
& \textit{``We're looking to retrofit into the existing structure. That's why we came up with the collaborative robot, because it's a robot that is small enough to fit in the current infrastructure.''}~(P8) \\
\addlinespace
\ref{sec:theme-testing} & Pilot tests 
& On-site pilot tests serve to validate technical feasibility and assess contextual fit. Innovation practitioners formulate ``risky assumptions'' to test but struggle to determine when technology is ``proven enough.'' Some question whether technology validation should be their core responsibility.
& \textit{``Those things you can't think of. You have to experience them in practice.''}~(P5) \\
\addlinespace
\ref{sec:theme-suppliers} & Technology suppliers 
& Suppliers are crucial in shaping projects; innovation practitioners conduct market scans and issue tender requests. A shift is emerging: from requesting ad-hoc solutions to keeping proposals open and adapting airport operations to fit commercially available technologies.
& \textit{``[We focus on] how surrounding operations could be modified, rather than on developing equipment fully fitted to their operations, which are anyway outdated and archaic.''}~(P4) \\
\addlinespace
\ref{sec:theme-human} & Human and operational aspects 
& Human and operational aspects are considered \textit{after} technology validation, through ``concepts of operation.'' Fallback situations are either deferred or linked to technology immaturity. Workers are positioned to support/supervise automation, requiring training and mindset shifts.
& \textit{``Once you've proven that the concept is feasible [...] then you could go to the stage of making it into an operational concept.''}~(P5) \\
\bottomrule
\end{tabularx}
\end{table*}
 

\subsection{Innovation Practitioners' Experiences with Automation Visions}
\label{sec:theme-visions}




\textcolor{black}{A vision is \textit{``just a starting point''} that gives direction to automation initiatives. Participants used the metaphor ``North Star''\footnote{In organizational strategy, a ``North Star'' refers to a guiding vision that provides consistent direction for decision-making, analogous to how the star Polaris has historically aided navigation.} to express that visions depict distant futures, yet provide guidance (\textit{``It's somewhere in the North, but where exactly we will end up? We don't know''}, P5). Currently, innovation practitioners follow the \textit{``Autonomous Airside Operations 2050''} vision, depicting the organization's ambition for a fully autonomous airside by 2050, as well as area-specific visions (e.g., \textit{Future baggage vision}). 
Visions identify potential automation projects by analyzing existing processes, proposing one-to-one replacements, and organizing them over time into complementary roadmaps. }

Visions are created mostly by innovation leads, often supported by external consulting companies. 
 In theory, visions should be dynamic and continuously developed, as P5 noted: \textit{``Its main purpose is to dictate direction, and then while developing towards such a goal, you will learn whether it is a good idea or not.''} However, iterations in practice remain unclear, and some innovation practitioners see visions as static. 

Innovation practitioners are assigned 
\textit{``specific initiatives within that entire scope''} (P5), which P2 saw as \textit{``top-down''}.
They engage in experimenting, researching, and scaling automation solutions, using automation concepts \textit{``as a vessel to learn, not per se as the end state''} (P5). 
Two main challenges emerged. First, translating visions to concrete projects is \textit{``tricky''} due to differences in scope and time-horizon;
P1 reflected that while the autonomous bus was tested only within specific routes and carrying airport personnel, the envisioned end goal would be to have a fleet service passengers throughout the entire airside.
Second, practitioners struggle to understand current milestones' value relative to the 2050 vision: \textit{``we just get projects, but we don't understand how that aligns with the bigger vision''} (P1), 
which requires ongoing conversations with innovation leads, as noted by P6: \textit{``everything that touches upon a vision you should do together with your innovation lead''}. Related to that, innovation practitioners sometimes experience \textit{``loneliness''}, since 
they carry significant project responsibility but face frustration when projects lack priority and resources. Some miss alignment between the autonomous airside vision and resource allocation. 

In summary, the iterative, guiding nature of visions contrasts with innovation practitioners’ perception of them as static, giving rise to translation challenges.

\subsection{The Reality of Project Development in Practice} 
\label{sec:theme-projects}

Automation projects not only originate in the vision, but can have less conventional starting points. Some projects are more opportunistic, spotted as \textit{``low hanging fruit''} (e.g., during student internships, site visits, fairs), while other projects come as requests from other departments: \textit{``Operations [department] started with the autonomous wheelchair project, and I was hooked to it to conduct final experimentation''} (P2). Innovation practitioners noted that it would be helpful to clarify the role of the innovation department, as they often take on projects other departments lack capacity for.


Projects can follow multiple 
aims, and 
participants highlighted the importance and effort of \textit{``protecting the scope.''} 
They explained that they easily \textit{``get sort of scope creep''} (P6), as colleagues from other departments often request testing additional features  within their pilot tests, or conducting demonstrations beyond the original project audience (e.g., \textit{``Hey, if you're testing an autonomous vehicle, can you also test a 5G network or a safety framework?''}, P6, referring to the autonomous baggage transportation project).
Innovation practitioners assess \textit{``the value of those side tracks''} 
 and their projects'  time and  scope, with special attention to how they manage suppliers' time; 
\textit{``I think that's important, because that's what your budgets are being signed off on.''} (P6)


This broader organizational context surrounding automation projects requires stakeholder management skills from innovation practitioners: \textit{``we often do projects not necessarily on the long-term agendas of everyone involved... So, people management, involving the right people, at the right time,  (...) I think is the majority of the work and really the hard part''} (P6). 
Innovation practitioners are sometimes required to produce alternative deliverables (e.g., business cases or communication plans), which also requires additional skills and flexibility. 
All in all, developing automation projects is messy and entangled rather than linear and isolated, and practitioners need to navigate organizational dynamics and the spontaneity of everyday work.





\subsection{Effect of Contextual Constraints in Solution Configurations} 
\label{sec:theme-constraints}

Since automation initiatives are often grounded in \textit{``specific demands from the business''} (P6), innovation practitioners begin by working closely with \textit{problem owners}\footnote{Term used by interviewees to refer to colleagues in the Operations Department who have an innovation problem to address}  to understand their challenges, identify \textit{``the gap''} to be addressed, and explore relevant use cases; P1, referring to the autonomous bus project, explained:  \textit{``I think that should be the starting point. So we have a shortage of drivers. So, when do we have a shortage of drivers? At peak times? When are the peak times?...''}.
These localized demands are then linked to broader airport challenges (e.g., staff shortages, increasing operational demand), enabling practitioners to build a value case that connects business needs, organizational priorities, and automation.

Therefore, a deep understanding of the context 
in which automation is to be deployed becomes central, and 
     shapes how automation is conceptualized.
First, solutions are often defined as a direct translation of the baseline manual operation, \textit{``[for the buses,] we were aiming for a one-on-one replacement''} (P1), where autonomous equipment is expected to be \textit{``technically capable''} of \textit{``mimicking''} (P6) a preliminary process.
Second,  solutions are strongly intertwined with or \textit{``dictated''} (P5) by contextual constraints, such as the operational rules and traffic culture of the airport, capacity limitations, or infrastructure configurations, including both physical assets and foundational digital technologies, 
\textit{``We're looking to retrofit into the existing infrastructure. That's why we came up with the collaborative [baggage handling] robot, because it's a robot that is small enough to fit''} (P8).
Innovation practitioners also noted the need to regulate the airport environment to favor the integration of autonomous assets.
Finally, several additional constraints were mentioned, including budget limitations, regulations, and supplier time zones (for those located in Asia or the Americas), that can affect the conceptualization stage:

\begin{quote}
    \textit{``Unfortunately, we're always in our constraints  [...] 
    it's like oh, but this would definitely not fit in my infrastructure, this is definitely way over budget, this might not have EU regulations, so that would take more time if it's a product coming in from the Americas or from Asia, then I need to get it EU regulated [...]
    So these are all things that we take into consideration''} (P4)
\end{quote}

Overall, the airport context significantly influences how automation is conceptualized, reflecting expectations to replicate airport processes, meet business demands, and navigate environmental constraints. The context provides innovation practitioners with a baseline for framing innovations and determining their specific features.

\subsection{Assessments of Technological Feasibility and Value through Pilot Tests}
\label{sec:theme-testing}


Once project scope is set and a preliminary automation concept defined, innovation practitioners conduct on-site experiments through a \textit{``proof of concept''} or \textit{``pilot test''}. This is the innovation department's main key performance indicator (KPI), as P1 explained, and involves lengthy testing to \textit{``mitigate risks''} (P2) before later stages. 

During pilot tests, the objective is twofold: validating technical feasibility and assessing contextual fit. First, innovation practitioners verify that solutions can technically mimic required behavior, according to specifications. The aim is to prove technology maturity for scaling:  \textit{``We want to make sure that the vehicle is technically capable of driving on an apron, connecting to a high loader, and picking up a container''} (P6, about the autonomous baggage transportation project). 
Second, practitioners assess automation's fit within airport infrastructure and operations. Referring to the autonomous bus project, P1 noted testing whether \textit{``an autonomous vehicle could adapt to the environment: can we meet the requirements of the operation? Can we move people from A to B in a safe way?''} They claim on-site testing is essential for understanding implementation consequences that suppliers cannot anticipate. In relation to the autonomous baggage transportation project, P5 explained:

\begin{quote}
    \textit{``When we did this trial with [supplier][...] it was driving in snow, but at a certain point it stopped not because of the snow, but because there are snow dunes by the side of the roads [...] So now we know our snow operations need to clear the road further than they do now. [...] A regular car can drive past that snow dune, it won't see it as an obstacle. So those things you can't think of. You have to experience them in practice.''} (P5)
\end{quote}

Innovation practitioners create lists of \textit{``risky assumptions''} to validate, often with stakeholders. 
The number of hypothesis formulated is usually high, so they also need to prioritize some of them.
At the end of the pilot, participants analyze the results and iterate on the initial automation concept.

However, determining when technology is \textit{``proven''} enough remains challenging---P3's \textit{``million-dollar question''}---as proof depends on predefined expectations that vary across stakeholders. While some accept imperfect solutions solving \textit{``even 50, 60, 70 or 80\%''} of problems as \textit{``good enough to start''} (P3), practitioners must often \textit{``convince''} others, especially given automation's long-term nature. 
For instance, the autonomous bus drove for ten consecutive months on a route, with some anomalies detected and solved, but this was not seen as all stakeholders as safe enough ---requiring practitioners to continue gathering evidence to convince the remaining skeptics. 
Similarly, in the collaborative baggage robot project, the robot proved it could handle baggage, but doing so sometimes more slowly than manual handling raised questions among some stakeholders about its practical usefulness.
Related to that, some practitioners critiqued the emphasis on technology validation, arguing that it is lengthy and outside their core expertise, and advocating a  focus on integration rather than technology maturity: 
    \textit{``We are an airport. We are not a technology developer [...] What we need to do is focus more on the second part of integration [...] bringing the system perspective in that integration part.''} (P1)

Overall, we observe that while on-site testing is essential for uncovering hard-to-anticipate implementation consequences, innovation practitioners struggle to proof technological feasibility, and even question whether this should be their responsibility.

\subsection{Market Offer and Collaborations with Technology Suppliers} 
\label{sec:theme-suppliers}

Automation solutions are sourced from the market, requiring innovation practitioners to stay aware of existing technologies through regular market scans across \textit{``all kinds of fields''} (P4) 
and to assess the maturity and scope of available solutions. 
This activity is sometimes outsourced to technology scouting experts, as was the case, for instance, in the autonomous ground power connection project, where suppliers in various robotics fields (e.g., medical, logistics) were assessed.

Suppliers are \textit{``a crucial aspect in the whole shaping of the project''} (P6), especially during piloting and scaling phases. 
Innovation practitioners define a minimum set of requirements 
as the basis for issuing open tenders.
From all contestants, a partner is selected for 
a defined pilot period, based on constraints and capabilities. Note that, occasionally, proofs of technology can also be scheduled outside this process when a strong supplier is identified directly. 
During the pilot, practitioners work closely with the selected supplier to set up systems both in the factory and on site, run experiments day-to-day, collect data, and manage other pilot-related activities.
Later, if a solution is validated and  the project receives green light to scale, 
practitioners return to the market to assess alternative suppliers and their specifications and development plans.

Participants reflected on their approach to suppliers. 
Some cases require creating solutions from scratch, because the technology \textit{``just does not exist out there''} (P4).
This was more common in the past, when suppliers saw little incentive to develop airport-specific products, given that Amsterdam Airport was the only potential use case.
In such cases, practitioners would modify existing solutions from different industries, as they did with the collaborative baggage handling robot, originally developed for food-processing use cases, adapting it to the airport context and \textit{``and eventually learn by doing that''} (P4). Today, with more companies on the market, they can leverage more tailored, context-specific products, such as the Foreign Object Debris removal vehicle or the autonomous wheelchair. Additionally, while in the past they would 
ask for ad hoc solutions from suppliers, they currently keep technology proposals very open, 
putting forth their requirements to the market yet acknowledging that operations can adapt to fit new solutions.

Given the crucial role of supplier relationships, we observe innovation practitioners becoming more open to adjusting airport operations to fit commercially available technologies and increasingly favoring collaboration over ad hoc solutions.

\subsection{Human Involvement and Operational Considerations}
\label{sec:theme-human}

After the proof of technology phase, innovation practitioners design a \textit{``concept of operation,''} which specifies the workers involved in a process, their roles, 
and how responsibilities are organized during anomalies and disruptions. As P5 noted, this is where they \textit{``make it really operational,''}
designing the process around automated equipment in detail. 
These operational aspects are deliberately deferred until technological feasibility is proven: 


\begin{quote}
    \textit{``Once you've proven that the concept is feasible, viable, and desirable, then you could go to the stage of making it into an operational concept. 
    But then we have already validated. We don't start with that because if the technology isn't the right fit, then you have a lot of detailed work with a lot of experts, and in the end, it doesn't work.
    ''
    }(P5)
\end{quote}

However, some participants reflected that process-related considerations are included \textit{``too late in the game''} (P1).
They 
would prefer an earlier involvement of the operations department, during solution conceptualizations, to \textit{``think along''} about the concept of operations.
Additionally, P1 and P7 noted that autonomous processes sometimes fail to address their intended problems and that reorganizing existing airport processes could be simpler. 


Regarding human involvement, while workers act as safety operators during pilot tests, practitioners expect to replace them with automation afterward. However, when asked about contingency plans, some  still expect humans to serve as a fallback: 
\textit{``26 employees are trained in the usage of the wheelchair and also in the events of malfunctions to reset the device or to do low-level resets''} (P2).
In contrast, others 
perceive fallback plans as unnecessary if the system is already proven: \textit{``I think if we feel this technology would not be proven enough, then you might need to have a fallback''} (P3).
Human considerations primarily concern new skills and roles, training workers to use autonomous technology, and managing their expectations and reactions regarding its adoption. As P4 shared: \textit{``for the collaborative robot, we're now trying to shift the mindset of people because we did face a lot of backlash on using it.''}

All in all, we observed that although human involvement and operational aspects inevitably influence the design and integration of solution setups, technology testing is prioritized over them.

\section{Findings (RQ2):  the Recursive Relationship between Automation Innovation Practices and Use Practices in Amsterdam Airport Schiphol} \label{sec: RQ2}

\begin{table*}[htb]
  \centering
  \caption{Comparison matrix between the themes of automation innovation practices in Amsterdam Airport Schiphol \& Co-performance principles. Summary: The co-performance principles \textit{``Design for recursive adaptation''} and \textit{``Design productive friction''} are in tension with the themes, while \textit{``Delegate according to capabilities''}, \textit{``Enable mutual responsiveness,''} and \textit{``Support co-learning over time''} are both in agreement (e.g., with contextual constraints and testing technology in context) and in tension (e.g., with operational aspects).} 

  \label{tab:co-performance-automation}
  \begin{tabularx}{\textwidth}{l l >{\raggedright\arraybackslash}X >{\raggedright\arraybackslash}X >{\raggedright\arraybackslash}X >{\raggedright\arraybackslash}X >{\raggedright\arraybackslash}X}
    \toprule
    & 
    \textbf{Theme{$\downarrow$}~Principle{$\rightarrow$}} & 
    \textbf{Design for recursive adaptation} &
    \textbf{Delegate according to capabilities} &
    \textbf{Enable mutual responsiveness} &
    \textbf{Design productive friction} &
    \textbf{Support co-learning over time} \\ 
    \cmidrule{3-7}
    \textbf{ID} & & \textbf{\texttt{A}} & \textbf{\texttt{B}} & \textbf{\texttt{C}} & \textbf{\texttt{D}} & \textbf{\texttt{E}} \\
    \midrule
    \textbf{\texttt{1}} & Vision as a starting point        & \ding{83} & \ding{81} & \ding{83} & ---       & \Circle  \\
    \textbf{\texttt{2}} & Project development in practice   & \ding{83} & ---       & \ding{83} & ---       & \ding{83} \\
    \textbf{\texttt{3}} & Contextual constraints            & ---       & \Circle   & \CIRCLE   & \ding{83} & ---       \\
    \textbf{\texttt{4}} & Testing technology in-context     & ---       & ---       & \CIRCLE   & \ding{83} & \Circle   \\
    \textbf{\texttt{5}} & Technology suppliers              & \ding{81} & ---       & \Circle   & ---       & ---       \\
    \textbf{\texttt{6}} & Operational aspects               & \ding{83} & \ding{81} & \ding{83} & ---       & \ding{83} \\
    \bottomrule
    \multicolumn{7}{l}{}\\
    \multicolumn{7}{l}{Legend: 
      \CIRCLE~\textit{strongly in agreement}; 
      \Circle~\textit{partially in agreement}; 
      ---~\textit{unrelated}; 
      \ding{83}~\textit{partially in tension};
      \ding{81}~\textit{strongly in tension}.
    } \\
  \end{tabularx}
\end{table*}

\textcolor{black}{In this section, we examine the relationship between automation innovation practices and the resulting use practices in our context, using the principles of co-performance drawn from the literature as our analytical lens (Table~\ref{tab:co-performance-principles}).
We examine these principles against the themes derived from our interviews (Section~\ref{sec:results}) to surface alignments and tensions, as summarized in Table~\ref{tab:co-performance-automation} \textcolor{black}{and further illustrated by Figure~\ref{fig:theoretical-model}}. 
This analysis identifies which dimensions of co-performance are supported by current innovation practices and which are in tension with them.
Co-performance provides us with a specific vocabulary to characterize this relationship, as well as the human–machine arrangements that emerge from it.}

\subsection{Prioritization of Fixed Arrangements over Situated Task Delegation}\label{sec:prioritization-of-fixed}

Innovation practitioners explained that their mindset is mostly dictated by organizational visions of automation, which depict fully autonomous airport processes, an \textit{``ambition''} they also follow in their concrete projects.
This may be in tension with the co-performance principle \textit{``Delegate according to capabilities''} (Table~\ref{tab:co-performance-automation}, cell 1B).
This principle distinguishes polymorphic actions (i.e., suitable for humans) from mimeomorphic actions (i.e., suitable for automation), emphasizing that each should perform the tasks they are best equipped for since their capabilities also \textit{evolve differently}.
In contrast, practitioners tend to view full automation as the end goal and aim to automate tasks whenever feasible, rather than distinguishing between task types.
This is also reflected in their preference for suppliers offering higher levels of autonomy in tender requests. 

Furthermore, in tension with the co-performance design principle \textit{``Enable mutual responsiveness''}, innovation practitioners mostly work toward fixed and pre-established worker-automation arrangements (1C and 6C). 
The principle suggests careful consideration of \textit{predetermined design decisions} to \textit{optimize for}, and rather establishes that \textit{appropriateness} emerges through situated practice.
This contrasts with \textit{``validating''} or \textit{``proving''}, terms practitioners used, that illustrate their view of automation as a closed goal to optimize.
By the time ground worker practices are considered, the type of automation system has typically already been decided upon, oriented toward full autonomy, and the possibilities for adapting them to fit their practices are severely limited.

As a result, even when innovation practitioners later recognized the limitations of these fixed, full-automation solutions ---once contextual factors (3B) or operational concepts (6B) were considered--- meaningful redesign was rarely possible. Instead, they relied on human workers to compensate for technological limitations and adapt solutions to the airport's operational environment.

\subsection{Lack of Openness, Responsiveness, and Reconfigurations in Innovation Processes}

The co-performance design principles \textit{``Design for recursive adaptation''} and \textit{``Enable mutual responsiveness''} suggest that \textit{dynamism} between humans and machines should be enabled and sustained over time.
In contrast, our context is largely characterized by \textit{single} automation configurations rather than changing worker-machine arrangements. 
As such, co-performance aspects related to \textit{responsiveness} and \textit{multiple forms of reconfiguration} (i.e., manual interventions, material settings adjustments, and system reprogramming) are largely limited to conceptualization and proof-of-technology stages and rather constrained once solutions become operational.

 \textit{``Design for recursive adaptation''} emphasizes that design continues \textit{through use practice}, enabling \textit{openness}.
In contrast, automation visions in our context tend to appear rigid and detached from operational realities, rather than flexible and iteratively reshaped through use practice (1A, 6A). 
Furthermore, the messy realities of innovation processes may be in tension with the co-performance aspect of openness (2A, 2C). 
Practitioners argued that the safety-critical and highly regulated nature of airports does not allow for spontaneous, dynamic adjustments in solution configurations. 
Additionally, innovation unfolds across multiple stakeholders and departments: 
innovation practitioners test solutions that are later scaled up by other departments, while end users belong to yet other departments, ultimatelly hindering openness.

We saw that innovation practitioners recognize workers as `co-designers,' in line with \textit{``Design for recursive adaptation''}, since their input and implication are being considered in some pilot tests.
Still, this is quite a top-down approach: more often, workers need to be \textit{``convinced''} about automation solutions, rather than considered as a co-designer. 

Finally, an additional tension affecting openness is the lack of in-house technological expertise, since most solutions are purchased from external suppliers (5A).
Yet, innovation practitioners reflected on the approaches they follow when contacting suppliers, and we saw a tendency towards more open-ended tender requests as well as a greater flexibility to adjust and renew airport operations around supplied solutions.


\subsection{The Tension between Productive Friction and Validation}\label{sec:friction-and-validation}

The co-performance design principle \textit{``Design productive friction''} is only weakly related to the themes that resulted from our interviews (4D and 5D).
This is the case because innovation practitioners focus on \textit{validating}, \textit{de-risking} and \textit{proving} that automated solutions are \textit{mature enough} to be transferred to a scale-up phase, which in their case translates into \textit{conflict-less}, \textit{friction-less}, or \textit{failure-less} solutions, in contrast to the notion of \textit{``appropriateness''}.
As such, they select suppliers demonstrating a certain level of maturity, subsequently define conditions 
of expected performance, test how well the solutions work in practice, monitor any anomalies (or \textit{frictions}) occurring when introducing autonomous equipment into the airport's operational environment, and refine the solutions to \textit{avoid} such frictions in the future.
While some innovation practitioners mentioned being more flexible toward imperfect automated solutions that could \textit{``only solve 50, 60, 70, or 80\% of a problem''} (P3), we observe an underlying assumption that systems should not fail once implemented in operations, or that failures would be something \textit{risky,} \textit{negative,} and \textit{unacceptable}.

In contrast, the principle sees everyday crises as opportunities for improved co-performances, and it suggests considering minor, productive frictions that enable learning without causing excessive distress as starting points for reconfigurations. 
It is necessary to create mechanisms for resolving conflicts about appropriate practice between human and artificial performers, 
since people differ, situations vary, and both people and situations change over time; therefore, a perfect solution is never possible. 


\subsection{Agreement on Situating Automation Concepts in Context}

The design principle \textit{``Enable mutual responsiveness''} emphasizes the idea that \textit{``appropriateness emerges through situated practice, rather than predetermined design decisions''},
which is strongly in agreement with the innovation practices of our participants, as they situate automation concepts in context when conceptualizing and validating them.
First, practitioners explained that the requirements for automation concepts are shaped by the airport's contextual limitations and specific features, as well as by existing regulations, organizational culture, or budget limitations.
Projects must align with business needs and operational rules, which innovation practitioners seek to \textit{``mimic''} or \textit{``not create conflict with''} (3C).

Second, innovation practitioners highly value on-site testing, mainly to assess technologies' fit within the airport context (4C). 
Some argued that this is the only way to understand the consequences and the side effects of its implementation in such a context, which is hard to replicate in lab environments.
Still, they also make special arrangements during testing, by placing autonomous equipment in safe, isolated airport areas, thereby removing the messiness of the context to some extent.
For instance, in the autonomous shuttle bus project, they started testing without involving passengers, but rather only allowing airport staff in, which might also affect appropriateness.

Finally, suppliers also recognize the need to modify their solutions based on the needs of the context, in line with the principle.
Since they are not used to operating in airport contexts and they lack knowledge about it, they acknowledge the need to run pilots on site and make adjustments in existing solutions accordingly (5C).

\subsection{Learning is Gradual and Limited to Proof of Technology Phases and Technology Development} 

The co-performance design principle \textit{``Support co-learning over time''} emphasizes that human and artificial performers need \textit{extended time} for \textit{mutual adaptation} and it accounts for the different temporal rhythms at which humans and machines learn.
This is partially in agreement with the automation visions of our context (1E), which portray \textit{gradual} \textit{transformations} over time in both the responsibilities assigned to machines and their level of automation in airport processes.
Similarly, proof of technology phases are intended to support  \textit{learning} during automation projects (4E), such as discovering pitfalls, solving them, and understanding both how solutions fit in context and how surrounding actors react.
However, these two aspects of innovation practice primarily emphasize technological development rather than mutual adaptation between workers and technology.
This is also reflected in the late consideration of human and operational aspects (6E), which typically only occurs after technology validation, requiring workers to adapt to already configured systems.

Furthermore, the principle is in tension with the practicalities of project development at the airport (2E). While it highlights the need for \textit{extended time to enact appropriate interfaces between humans and technologies}, these learnings are temporally confined to the proof of technology phase rather than continuing into later deployment, scaling, and use stages.

\begin{figure*}[htbp]
  \centering
  \includegraphics[width=1.0\textwidth]{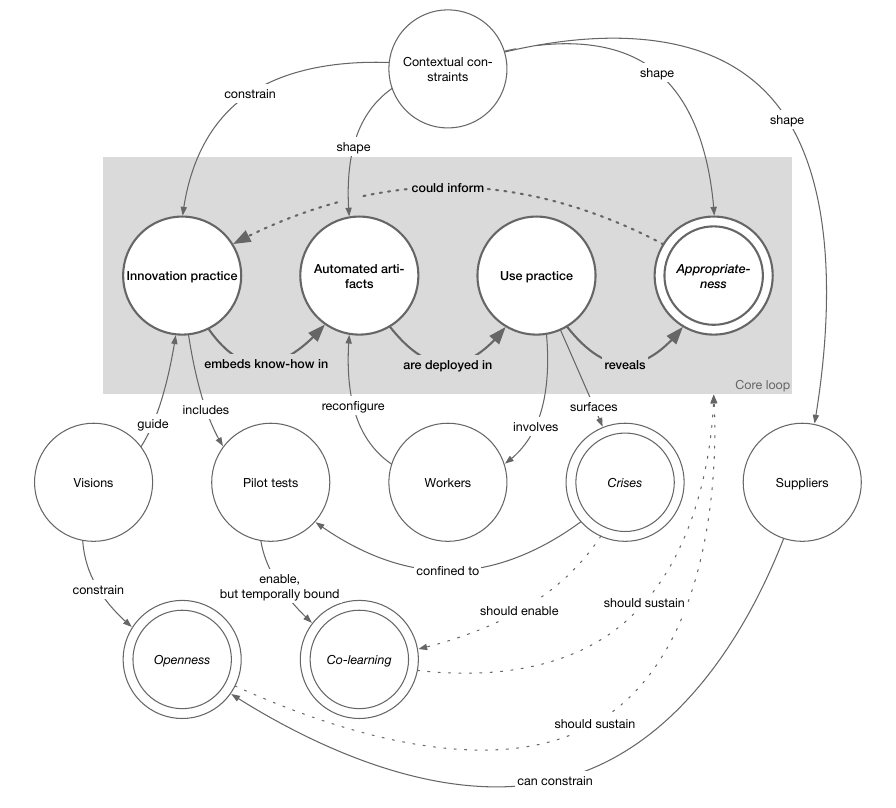}
  \caption{Conceptual model of the recursive relationship between innovation practice and use practice, synthesized from our findings and interpreted through co-performance theory. Double-border nodes represent co-performance concepts (Appropriateness, Openness, Co-learning, Crises); single-border nodes represent empirical elements. The core loop captures how practitioners embed know-how in automated artifacts, deployed in use practice, revealing appropriateness. Co-performance theory does not explicitly articulate a feedback loop from appropriateness back to professional innovation practice; the ``could inform'' edge (dashed) represents an underexplored connection that would enable practitioners to incorporate knowledge from use into subsequent cycles (Section~\ref{disc: IMPLICATIONS}). Contextual constraints shape both artifacts and appropriateness criteria. Two co-performance enablers---openness and co-learning---should sustain the recursive loop but are blocked: visions constrain openness when treated as fixed end-states (Section~\ref{sec:theme-visions}), and pilot tests temporally bound co-learning (Section~\ref{sec:theme-testing}). Crises surface through use practice and should enable co-learning, but are confined to pilot tests rather than treated as ongoing learning opportunities (Section~\ref{sec:friction-and-validation}). Workers provide an additional recursive pathway through artifact reconfiguration---underutilized in current practice (Section~\ref{sec:theme-human}). Dotted ``should'' edges indicate co-performance prescriptions that our analysis reveals are weakened in this context. This visualization foregrounds key dynamics; additional relations are omitted for clarity.}
  \label{fig:theoretical-model}
  \Description{A concept map showing the recursive relationship between innovation practice and use practice in workplace automation, derived from interview data at an international airport. The diagram contains twelve nodes connected by labeled directional edges; four nodes representing co-performance concepts (Appropriateness, Openness, Co-learning, Crises) are distinguished by double borders, while eight nodes representing empirical elements have single borders. Four nodes form a central loop: Innovation Practice connects to Automated Artifacts via ``embeds know-how in''; Automated Artifacts connects to Use Practice via ``are deployed in''; Use Practice connects to Appropriateness via ``reveals''; and Appropriateness connects back to Innovation Practice via ``could inform'' (dashed line), completing the loop. Outside this loop, Contextual Constraints has three outgoing edges: ``shape'' pointing to Automated Artifacts, ``shape'' pointing to Appropriateness, and ``constrain'' pointing to Innovation Practice. Visions connects to Innovation Practice via ``guide'' and to Openness via ``constrain.'' Openness connects to the recursive loop via ``should sustain'' (dotted line). Pilot Tests is connected from Innovation Practice via ``includes'' and connects to Co-learning via ``enable, but temporally bound.'' Co-learning connects to the recursive loop via ``should sustain'' (dotted line). Crises is connected from Use Practice via ``surfaces,'' connects to Pilot Tests via ``confined to,'' and connects to Co-learning via ``should enable'' (dotted line). Workers connects from Use Practice via ``involves'' and back to Automated Artifacts via ``reconfigure,'' forming a secondary recursive pathway that the study found to be underutilized in current practice. Suppliers connects to Openness via ``can constrain.''}
\end{figure*}

\section{Discussion}\label{sec:discussion}



The purpose of this paper was to scrutinize automation innovation practices at Amsterdam Airport Schiphol, to analyze how those practices affect the conceptualization of worker-automation arrangements. 
Our empirical findings illustrate innovation practitioners' tasks, mindset, development processes, stakeholder collaborations, and tests (RQ1, Section~\ref{sec:results}). 
Using co-performance as a theoretical lens, we revisit these themes to capture the recursive relationship between professional innovation and use practices of automated artifacts and examine what ideas regarding appropriate performance are materialized through them (RQ2, Section \ref{sec: RQ2}).
Based on these findings, in this section, we offer implications for practice and future research (\ref{disc: IMPLICATIONS}) and reflect on the use of the theoretical lens of co-performance in organizational contexts (\ref{disc: co-perf}).

\subsection{Implications for Practice and Research}
\label{disc: IMPLICATIONS}

The following implications for HCI and design research and practice discuss how future work might rethink and guide automation innovation practices in organizations in ways that enable appropriate worker-automation co-performances.

\subsubsection{Including dynamic task delegation and situated reconfigurations in automation visions} 
\label{discussion: vision}



While full automation visions may be intentionally aspirational \cite{Simonse2024},  they shape shared understandings and imaginaries of automation \cite{GomezBeldarrain2025, Baur2023, Martin2021} and practitioners are often unaware of their unrealistic nature~\cite{Bradshaw2013, Riek2025}. 
In our context, when the 2050 Automation vision was  translated into a concrete, horizon-based roadmap, these same issues resurfaced across the different timescales that make up the 30-year automation program.
Consequently, full automation goals resulted in practitioners considering workers only at later stages, leaving humans with residual tasks or responsibility for integrating technology into broader process pipelines~\cite{Fox2023, Delfanti2021, MontielValle2024}. 
These expectations, that humans  \textit{extend} and \textit{sustain} machinery \cite{Delfanti2021}, should not be undervalued as new forms of labor~\cite{Ma2025}. 
We argue that supporting co-performance requires reframing visions, to emphasize dynamic worker-automation relationships rather than fixed, full automation arrangements. 

\hfill

\textbf{Implications for practice:} 
Innovation teams should enable some level of iteration in automation visions over time, incorporating new knowledge that practitioners acquire through context, actual use, and situated practice.
In that regard, special attention should be given to  \textit{futures literacy} and \textit{anticipatory capacities} of practitioners, terms that describe ``openness to possible futures and sensitivity to possibilities, as opposed to seeking accurate probability-oriented
prediction'' \cite[p. 389]{Epp2022}.  
As illustrated in Figure~\ref{fig:theoretical-model}, establishing a feedback loop between situated appropriateness and professional innovation practice is a concrete first step organizations can take. 

\textbf{Implications for research:}  A longer-term research agenda could ask what it would mean 
to incorporate openness into automation visions, rather than merely updating them periodically.
Building on recent HCI futures studies \cite{Sanchez2025, Salovaara2025, BENDOR2021}, particularly those studying innovation teams \cite{Epp2022, Salovaara2022}, future research could explore 
\textit{living visions} that evolve through practice. 
This opens broader questions: who gets to revise a vision, at what moments, and through what mechanisms? 
HCI can guide practitioners and organizations in building visions that 
allow configurations to emerge and adapt over time.

\subsubsection{Reframing organization-supplier relationships to enable solution reconfigurations during use} \label{discussion: suppliers}
Practices for sourcing, testing, adapting, and integrating supplier solutions are often in tension with aspects of co-performance and
the lack of in-house technological expertise makes the airport dependent on locked-in vendor solutions, hindering \textit{openness} and \textit{dynamic rearrangements}. 
Still, since most suppliers lack expertise in airport contexts,  they acknowledge the need to run pilot tests on-site and adapt their solutions to local requirements, in line with \textit{``enable mutual responsiveness''}.

Related to that, we argue that \textit{reconfigurations} of supplier solutions during use, by \textit{everyday designers} (i.e., workers), should be further enabled.
Rather than relying on fixed specifications that constrain solutions from the outset, tender processes and contracts should also support situated learning---allowing supplier technologies to evolve through real-world testing and enabling multiple pathways for reconfiguration as context reveals what actually works. 
This would resonate with the ``adapt'' phase for innovation-adopting organizations in \citet{DAMANPOUR2006_inn}. \\ 

\textbf{Implications for practice:}
Practitioners should remain aware that external solutions should not be viewed as fully finished.
Organizations should anticipate the time and effort required for \textit{reconfiguration,}   negotiate reconfiguration rights with suppliers, and consistently include workers in the process---from early supplier consideration through post-deployment adjustments.
For instance, innovation practitioners could involve ground workers when weighing multiple supplier options.

\textbf{Implications for research:} 
Our study raises questions about how to facilitate more open-ended tender processes, in which solutions are co-developed by suppliers and organizations to better adapt to local contexts, rather than fixating on closed solutions that limit reconfigurations. 
Future research could develop guidance for practitioners  sourcing automation technologies, for example through guidelines,  checklists, or model contracts that support adaptation (e.g., \cite{Alfrink_IASDR23}).
    
\subsubsection{The rigidity of airport contexts might constrain appropriate worker–automation arrangements}
\label{discussion: concepts}

Innovation  practices at Amsterdam Airport Schiphol are aligned with the co-performance principle \textit{``enable mutual responsiveness''}, as they account for contextual situatedness to enable appropriateness.
However, the rigidity of the airport context might be constraining, 
as airport operations are hard to change, making \textit{mutual adaptation} difficult.
Moreover, automation solutions are often defined as direct  translations of isolated manual operations, aiming to “mimic” existing contexts rather than enabling reconfigurations in surrounding operational requirements, human aspects, and  scenarios. 
Such ``reductionist'' conceptualizations of automation have been widely studied \cite{Fox2023, Delfanti2021}, and are often a consequence of myths of autonomy \cite{Bradshaw2013, Riek2025} and understanding automation as a ``simple substitute of human capability'' \cite{Bradshaw2013}.

\hfill

\textbf{Implications for practice:} 
Automation concepts should be defined by simultaneously considering technological, operational, and human aspects, rather than as direct translations of isolated manual tasks. 
This requires integrating operational and worker expertise early in the design process, to account for real-world dynamism and identify which contextual constraints have flexibility for change, how operations are expected to evolve, and what are the priorities of the involved stakeholders. 
Innovation practitioners should capture this information systematically, for instance through conversations and structured workshops with multiple stakeholders and visualizations of how operations are expected to evolve.

\textbf{Implications for research:} 
HCI research should continue to challenge myths about automation technologies and provide clear guidance to practitioners, consistent with prior work \cite{Baur2023, Riek2025, GomezBeldarrain2025}. 
Besides, future research could support the design of automation concepts as living documents that evolve with changing contextual constraints, never tethered to frozen operational models, and always incorporating fallback scenarios for when things go wrong.


\subsubsection{Friction and crises, beyond proof of technology stages, are challenging in safety-critical settings} 
\label{discussion: proof of tech} 



The use of \textit{friction} (e.g. intentional slowness) in design is not a novel concept \cite{Gould2021, Haliburton2024, Garret2025, van_beek_everyday_2025}; however, in co-performance it mainly involves considering mismatches in the appropriateness of a given co-performance as a starting point and an opportunity for reconfiguration.
In Amsterdam Airport Schiphol, friction is limited to testing phases only. 
This might be the case because  co-performance theories are mostly drawn from domestic contexts, where a failure has fewer cascading effects than in our context, which must ensure operational continuity and could be heavily disrupted, as highlighted in safety-critical automation domains~\cite{Rozzi2012, Amaldi2012, Perry2005, Demirci2021}.\\

\textbf{Implications for practice:} 
Organizations should establish protocols that treat crises and breakdowns in automated operations as decision points for reconfiguration rather than automatic returns to prior arrangements---for instance, requiring a joint review with workers before reinstating a new setup.
Beyond crisis moments, since friction also manifests as smaller, everyday mismatches that workers notice and resolve informally, organizations should create a channel (e.g., a recurring debrief, a shared log) through which workers can flag these mismatches so they inform ongoing automation refinement.


\textbf{Implications for research:} 
HCI and design researchers could think about how to consider crisis and frictions during the use of automation, to enable learning without compromising on the safety-critical values that are relevant for such contexts, as well as how to resolve conflicts about appropriate co-performances beyond proof of technology stages. 

\subsubsection{Transforming pilot tests from technology validation into collaborative experimentation spaces}
\label{discussion: proof of tech}

Our participants view on-site testing as a way to understand the consequences and side effects of automated technology within the airport context.
During these tests, practitioners focus on \textit{de-risking} and \textit{optimizing} the technology against factors and hypotheses that they establish beforehand. 
As such, the desired outcome 
is the \textit{``validation''} of solutions, that is, ensuring that technology is sufficiently mature to perform specific tasks without failure.
In contrast, co-performance rather highlights that design is recursive, with humans and technologies learning and gradually transforming their attitudes to enact appropriate interfaces.
Moreover, tests should not be conceived as time-constrained units, but rather as ongoing processes that continue through use practice.\\


\textbf{Implications for practice:} Organizations should reframe pilot tests as opportunities for \textit{situated learning}, where the outcome is not validation but rather new knowledge of how automated solutions perform within specific operational contexts. 
Since humans and technologies learn at different paces and require extended time to learn to work together, innovation timelines should account for this gradual, recursive process.


\textbf{Implications for research:} 
Future research could transform pilot tests  from technology validation exercises into collaborative discovery spaces. 
Those tests could reveal how workers and machines learn to perform together---embracing productive friction and real-world messiness while recognizing workers as co-designers who actively shape automation through their everyday adaptations.
To support innovation practitioners in building such collaborative discovery spaces, HCI research could design specific study protocols and spaces. While methods could be borrowed from existing research \cite{van_beek_everyday_2025, Rozendaal2024}, the main challenge lies in adapting them to operational, safety-critical contexts, where experimentation and enactment must co-exist with established protocols and real-world circumstances.

\subsubsection{Enabling openness, adjustments, and learning through departments and project pipelines}
\label{discussion: concepts}

Prior work recognizes the gap between innovation pilots and corporate integration~\cite{Klitsie2019, hellstrom2002}, noting the impact of departmental silos and divergent innovation priorities~\cite{Klitsie2019}.
In our context, innovation unfolds across multiple departments, requiring handovers and coordination between many stakeholders: while innovation practitioners conceptualize and test solutions, these are later scaled by other departments, and end users belong yet to other departments. This ultimately hinders \textit{openness}, as shown by our findings. \\

\textbf{Implications for practice:} 
Anticipating the complexity of innovation pipelines, organizations should establish  procedures that allow for iteration not only during testing, but throughout all stages ~\cite{hellstrom2002}, including deployment and use~\cite{GYLDENKAERNE2024, Zajac2025, Fox2024_Humane}. 
This requires structured handover processes between departments to preserve knowledge across stages, as well as mechanisms for retaining insights from each project to inform future ones \cite{Freire2023}. 
Additionally,  multi-stakeholder working groups  should oversee projects across their full life cycle. Without such mechanisms, departmental silos will continue to constrain the adaptability that co-performance requires.


\textbf{Implications for research:} 
To break the illusion of linear innovation pipelines, HCI research should support practitioners in the messy organizational realities through which automation unfolds---crossing departments, shifting ownership, and evolving unpredictably. Future research could build mechanisms and procedures that capture and share learnings across these boundaries 
and allow other parties to iterate preliminary worker-automation arrangements.\\

    
\subsection{Reflections on Applying Co-Performance to Organizational Innovation Contexts} 
\label{disc: co-perf}

This paper brings a different orientation to the use of co-performance theory to examine appropriateness in human-automation arrangements.
While prior work \cite{van_beek_everyday_2025, van_beek_making_2023, Viaene2021, kuijer_automated_2019} mostly focuses on domestic  practices and use time, where residents act as recipients of technology, in our case we use the theory to understand work practices and design time, where innovation practitioners act as developers of technology.
Hence, these differences in focus bring novel insights into what matters in the recursive relationship between professional design practices and use practices of automated technologies when transferred into organizational innovation contexts (cf. Figure~\ref{fig:theoretical-model}).

While domestic practices (i.e., ``everyday practices,'' \cite{kuijer_automated_2019}) are social practices shaped by socio-cultural norms, work practices differ in that they fall within organizational practices. 
While not exempt from cultural norms, work practices entail greater complexity: they are strictly dictated by organizations, governed by regulations, and professionally enacted by workers. They require specific skills and are embedded in contractual relations.  
Organizational complexity and strict regulations introduce constraints that differ substantially from the everyday social norms shaping domestic practices.
For instance, in our airport context, workers must adhere to strict timing constraints and perform their tasks in prescribed ways to protect themselves, passengers, airside workers, and equipment.

However, organizations entail closer feedback loops with workers, as innovation and use practices are much more closely intertwined than in home environments.
Similarly, organizations may engage directly with suppliers at various stages and in different settings, thereby enabling feedback loops that inform automation development.
These two aspects make it easier to create \textit{collaborative experimentation spaces} that could potentially reveal how workers and machines perform together,  recognizing workers as co-designers who actively shape automation through their everyday adaptations, as discussed in this paper.

Further, we recognize that in organizational contexts, design decisions do not rest solely with innovation practitioners; they are also shaped by technology suppliers, authorities, and other stakeholders in the same context, all of whom exert  significant influence over the technologies that ultimately reach workers. 
A more nuanced distinction is needed between human actors who influence performances through the know-how they embed in intelligent machines, referred to by \citet{kuijer_co-performance_2018} as `designers'.

\subsection{Limitations and Future Work}


Having conducted interviews within a single innovation-adopting organization---an international airport in Western Europe---the resulting findings reflect specific innovation practices, company culture, team dynamics, use practices, and constraints from the aviation sector. 
Hence, the primary limitation of our study concerns transferability. 
Readers should carefully consider which of the discussed agreements and tensions around co-performance apply to their context.

This situatedness was deliberate. 
We aimed to provide thick, in-depth descriptions of practitioner challenges, where organizational intricacies and contextual realities critically shape automation innovation as well as its recursive relationship with use practices. 
Future research could replicate our methodology and extend these learnings in alternative organizational settings---such as, manufacturing, healthcare, logistics---to identify which insights generalize across contexts versus which remain case-specific.
In that regard, we aimed to offer details that would allow others to follow a similar analytic procedure.

Our implications for HCI and design research aim to inform future work on automation innovation practices and motivate the development of targeted interventions to support practitioners.
For example, tools that support iterative vision development (\ref{discussion: vision}), model contracts to ease adaptation of supplier solutions (\ref{discussion: suppliers}), or study protocols to allow experimentation between workers and automation in safety-critical contexts (\ref{discussion: proof of tech}). 
Additionally, future research must acknowledge the messiness of everyday organizational practices when studying workplace automation. 
HCI's tools and guidelines will only be useful if they account for the realities of scope creep, stakeholder tensions, and the gap between innovation pilots and operational integration that practitioners navigate daily \cite[cf.][]{dourish_where_2004, suchman_human-machines_2009}. 


\section{Conclusion}\label{sec:conclusion}

This study analyzes end-to-end automation innovation practices in a Western European airport, 
through a co-performance lens.
The aim is to foreground the recursive relation between professional innovation practice and use practice in automation projects, to rethink prevailing techno-optimistic approaches that sideline human workers and lead to failed technology adoption.
We find that full-automation narratives in company visions lead innovation practitioners to prioritize fixed worker–automation arrangements and delay task delegation.
Furthermore, while practitioners successfully rely on contextual inputs and validations to achieve appropriateness, the analyzed projects lack openness  and therefore fail to enable workers to be responsive and to reconfigure automation solutions during later use.
As an initial step toward addressing these issues, we reflect on the implications of a co-performance lens to reframe innovation approaches in organizations,
offering insights into specific innovation practices which shape, but rarely appear in, HCI's automation discourse.

\begin{acks}
This research was funded by a Public-Private Partnership for Research and Development (PPP allowance) from the Dutch Ministry of Economic Affairs and Climate Policy via Click NL and the Royal Schiphol Group.

The authors are deeply grateful to all participants for their time and valuable contributions, and to the anonymous reviewers for their constructive feedback, which greatly improved the manuscript.
Finally, we thank Rosina Kotey for her support during the research and Mireia Yurrita Semperena, Katherine Song, and Céline Offerman for their feedback on the various versions of this manuscript.
 
\end{acks}

\bibliographystyle{ACM-Reference-Format}
\bibliography{Sections/references}

@proceedings{Joo2024,
    author = {Joo, J. and Gomez-Beldarrain, G. and Kim, E. and Verma, H.},
    title = {Transition towards automatic Passenger Boarding Bridge: Themes of task delegation for autonomous airport operations},
    year = {2024},
    booktitle = {in Gray, C., Ciliotta Chehade, E., Hekkert, P., Forlano, L., Ciuccarelli, P., Lloyd, P. (eds.), DRS2024: Boston, 23–28 June, Boston, USA}, 
    doi ={https://doi.org/10.21606/drs.2024.872}
}

@inproceedings{kuutti_turn_2014,
  title = {The Turn to Practice in {{HCI}}: Towards a Research Agenda},
  shorttitle = {The Turn to Practice in {{HCI}}},
  booktitle = {Proceedings of the {{SIGCHI Conference}} on {{Human Factors}} in {{Computing Systems}}},
  author = {Kuutti, Kari and Bannon, Liam J.},
  year = 2014,
  month = apr,
  series = {{{CHI}} '14},
  pages = {3543--3552},
  publisher = {Association for Computing Machinery},
  address = {New York, NY, USA},
  doi = {10/gf65fj},
  urldate = {2025-11-07},
  isbn = {978-1-4503-2473-1},
}

@inproceedings{amershi_guidelines_2019,
  title = {Guidelines for {{Human-AI Interaction}}},
  booktitle = {Proceedings of the 2019 {{CHI Conference}} on {{Human Factors}} in {{Computing Systems}}},
  author = {Amershi, Saleema and Weld, Dan and Vorvoreanu, Mihaela and Fourney, Adam and Nushi, Besmira and Collisson, Penny and Suh, Jina and Iqbal, Shamsi and Bennett, Paul N. and Inkpen, Kori and Teevan, Jaime and {Kikin-Gil}, Ruth and Horvitz, Eric},
  year = 2019,
  month = may,
  series = {{{CHI}} '19},
  pages = {1--13},
  publisher = {Association for Computing Machinery},
  address = {New York, NY, USA},
  doi = {10/ggbzdg},
  urldate = {2025-12-01},
  isbn = {978-1-4503-5970-2},
}

@article{gombolay_computational_2017,
  title = {Computational Design of Mixed-Initiative Human--Robot Teaming That Considers Human Factors: Situational Awareness, Workload, and Workflow Preferences},
  shorttitle = {Computational Design of Mixed-Initiative Human--Robot Teaming That Considers Human Factors},
  author = {Gombolay, Matthew and Bair, Anna and Huang, Cindy and Shah, Julie},
  year = 2017,
  month = jun,
  journal = {The International Journal of Robotics Research},
  volume = {36},
  number = {5-7},
  pages = {597--617},
  publisher = {SAGE Publications Ltd STM},
  issn = {0278-3649},
  doi = {10/gbf88x},
  urldate = {2025-12-01},
  langid = {english},
}

@article{hollan_distributed_2000,
  title = {Distributed Cognition: Toward a New Foundation for Human-Computer Interaction Research},
  shorttitle = {Distributed Cognition},
  author = {Hollan, James and Hutchins, Edwin and Kirsh, David},
  year = 2000,
  month = jun,
  journal = {ACM Trans. Comput.-Hum. Interact.},
  volume = {7},
  number = {2},
  pages = {174--196},
  issn = {1073-0516},
  doi = {10/cbdnrv},
  urldate = {2025-12-01},
}

@article{kaber_effects_2004,
  title = {The Effects of Level of Automation and Adaptive Automation on Human Performance, Situation Awareness and Workload in a Dynamic Control Task},
  author = {Kaber, David B. and Endsley, Mica R.},
  year = 2004,
  month = mar,
  journal = {Theoretical Issues in Ergonomics Science},
  volume = {5},
  number = {2},
  pages = {113--153},
  publisher = {Taylor \& Francis},
  issn = {1463-922X},
  doi = {10/dnmhc2},
  urldate = {2025-12-01},
}

@article{lee_trust_2004,
  title = {Trust in {{Automation}}: {{Designing}} for {{Appropriate Reliance}}},
  shorttitle = {Trust in {{Automation}}},
  author = {Lee, John D. and See, Katrina A.},
  year = 2004,
  month = mar,
  journal = {Human Factors},
  volume = {46},
  number = {1},
  pages = {50--80},
  publisher = {SAGE Publications Inc},
  issn = {0018-7208},
  doi = {10/dr6jf9},
  urldate = {2025-12-01},
  langid = {english},
}

@article{parasuraman_model_2000,
  title = {A Model for Types and Levels of Human Interaction with Automation},
  author = {Parasuraman, R. and Sheridan, T.B. and Wickens, C.D.},
  year = 2000,
  month = may,
  journal = {IEEE Transactions on Systems, Man, and Cybernetics - Part A: Systems and Humans},
  volume = {30},
  number = {3},
  pages = {286--297},
  issn = {1558-2426},
  doi = {10/c6zf92},
  urldate = {2025-12-01},
}

@article{rogers_distributed_1994,
  title = {Distributed {{Cognition}}: {{An Alternative Framework}} for {{Analysing}} and {{Explaining Collaborative Working}}},
  shorttitle = {Distributed {{Cognition}}},
  author = {Rogers, Yvonne and Ellis, Judi},
  year = 1994,
  month = jun,
  journal = {Journal of Information Technology},
  volume = {9},
  number = {2},
  pages = {119--128},
  publisher = {SAGE Publications Ltd},
  issn = {0268-3962},
  doi = {10/gpckjt},
  urldate = {2025-12-01},
  langid = {english},
}

@book{BraunC,
  title={Thematic analysis, a practical guide},
  author={Virginia Braun and Victoria Clarke},
  year={2021},
  publisher={Sage publications}
}

@article{Bradshaw2013,
   author = {Jeffrey M. Bradshaw and Robert R. Hoffman and David D. Woods and Matthew Johnson},
   doi = {10.1109/MIS.2013.70},
   issn = {15411672},
   issue = {3},
   journal = {IEEE Intelligent Systems},
   pages = {54-61},
   title = {The seven deadly myths of 'autonomous systems'},
   volume = {28},
   year = {2013},
}

@inproceedings{Baldauf2021ws,
author = {Baldauf, Matthias and Fr\"{o}hlich, Peter and Sadeghian, Shadan and Palanque, Philippe and Roto, Virpi and Ju, Wendy and Baillie, Lynne and Tscheligi, Manfred},
title = {Automation Experience at the Workplace},
year = {2021},
isbn = {9781450380959},
publisher = {Association for Computing Machinery},
address = {New York, NY, USA},
url = {https://doi.org/10.1145/3411763.3441332},
doi = {10.1145/3411763.3441332},
booktitle = {Extended Abstracts of the 2021 CHI Conference on Human Factors in Computing Systems},
articleno = {89},
numpages = {6},
location = {Yokohama, Japan},
series = {CHI EA '21}
}

@article{Wang2019,
author = {Wang, Dakuo and Weisz, Justin D. and Muller, Michael and Ram, Parikshit and Geyer, Werner and Dugan, Casey and Tausczik, Yla and Samulowitz, Horst and Gray, Alexander},
title = {Human-AI Collaboration in Data Science: Exploring Data Scientists' Perceptions of Automated AI},
year = {2019},
issue_date = {November 2019},
publisher = {Association for Computing Machinery},
address = {New York, NY, USA},
volume = {3},
number = {CSCW},
url = {https://doi.org/10.1145/3359313},
doi = {10.1145/3359313},
journal = {Proc. ACM Hum.-Comput. Interact.},
month = {nov},
articleno = {211},
numpages = {24}
}

@article{Breuer2023,
author = {Breuer, Svenja and Braun, Maximilian and Tigard, Daniel and Buyx, Alena and M\"{u}ller, Ruth},
title = {How Engineers’ Imaginaries of Healthcare Shape Design and User Engagement: A Case Study of a Robotics Initiative for Geriatric Healthcare AI Applications},
year = {2023},
issue_date = {April 2023},
publisher = {Association for Computing Machinery},
address = {New York, NY, USA},
volume = {30},
number = {2},
issn = {1073-0516},
url = {https://doi.org/10.1145/3577010},
doi = {10.1145/3577010},
journal = {ACM Trans. Comput.-Hum. Interact.},
month = {mar},
articleno = {30},
numpages = {33}
}

@article{GYLDENKAERNE2024,
title = {Innovation tactics for implementing an ML application in healthcare: A long and winding road},
journal = {International Journal of Human-Computer Studies},
volume = {181},
pages = {103162},
year = {2024},
issn = {1071-5819},
doi = {https://doi.org/10.1016/j.ijhcs.2023.103162},
url = {https://www.sciencedirect.com/science/article/pii/S1071581923001714},
author = {Christopher Gyldenkærne and Jens Ulrik Hansen and Morten Hertzum and Troels Mønsted},
}

@inproceedings{Rozzi2012,
author = {Rozzi, Simone and Amaldi, Paola},
year = {2012},
month = {05},
pages = {98-106},
title = {Organizational and inter-organizational precursors to problematic automation in safety critical domains},
publisher = {Association for Computing Machinery},
booktitle = {ATACCS '12: Proceedings of the 2nd International Conference on Application and Theory of Automation in Command and Control Systems},
address = {New York, NY, USA},
}

@inproceedings{Eisser2020,
author = {Ei\ss{}er, Judith and Torrini, Mario and B\"{o}hm, Stephan},
title = {Automation Anxiety as a Barrier to Workplace Automation: An Empirical Analysis of the Example of Recruiting Chatbots in Germany},
year = {2020},
isbn = {9781450371308},
publisher = {Association for Computing Machinery},
address = {New York, NY, USA},
url = {https://doi.org/10.1145/3378539.3393866},
doi = {10.1145/3378539.3393866},
booktitle = {Proceedings of the 2020 on Computers and People Research Conference},
pages = {47–51},
numpages = {5},
location = {Nuremberg, Germany},
series = {SIGMIS-CPR'20}
}

@inproceedings{Mlynar2022,
author = {Mlynar, Jakub and Bahrami, Farzaneh and Ourednik, Andr\'{e} and Mutzner, Nico and Verma, Himanshu and Alavi, Hamed},
title = {AI beyond Deus ex Machina – Reimagining Intelligence in Future Cities with Urban Experts},
year = {2022},
isbn = {9781450391573},
publisher = {Association for Computing Machinery},
address = {New York, NY, USA},
url = {https://doi.org/10.1145/3491102.3517502},
doi = {10.1145/3491102.3517502},
booktitle = {Proceedings of the 2022 CHI Conference on Human Factors in Computing Systems},
articleno = {370},
numpages = {13},
location = {New Orleans, LA, USA},
series = {CHI '22}
}

@article{bainbridge1983,
 author = {L Bainbridge},
   title = {Ironies of Automation},
   year = {1983},
    journal = {Automatica},
   pages = {775-779},
    volume = {19},
     issue = {6},
   publisher = {ACM},
}

@inproceedings{Amaldi2012,
author = {Amaldi, Paola and Smoker, Anthony},
title = {The problem with automation is not over-automation but lack of automation policy},
year = {2012},
isbn = {9782917490204},
publisher = {IRIT Press},
address = {Toulouse, FRA},
booktitle = {Proceedings of the 2nd International Conference on Application and Theory of Automation in Command and Control Systems},
pages = {176–181},
numpages = {6},
location = {London, United Kingdom},
series = {ATACCS '12}
}

@article{Fox2023,
author = {Fox, Sarah E. and Shorey, Samantha and Kang, Esther Y. and Montiel Valle, Dominique and Rodriguez, Estefania},
title = {Patchwork: The Hidden, Human Labor of AI Integration within Essential Work},
year = {2023},
issue_date = {April 2023},
publisher = {Association for Computing Machinery},
address = {New York, NY, USA},
volume = {7},
number = {CSCW1},
url = {https://doi.org/10.1145/3579514},
doi = {10.1145/3579514},
journal = {Proc. ACM Hum.-Comput. Interact.},
month = {apr},
articleno = {81},
numpages = {20}
}

@article{IVANOV2020,
title = {Automation fears: Drivers and solutions},
journal = {Technology in Society},
volume = {63},
pages = {101431},
year = {2020},
issn = {0160-791X},
doi = {https://doi.org/10.1016/j.techsoc.2020.101431},
url = {https://www.sciencedirect.com/science/article/pii/S0160791X20300488},
author = {Stanislav Ivanov and Mihail Kuyumdzhiev and Craig Webster},
}

@article{Baur2023,
title = {Inserting machines, displacing people: how automation imaginaries for agriculture promise ‘liberation’ from the industrialized farm.},
journal = {Agriculture and Human Values},
volume = {40},
pages = {815-833},
year = {2023},
doi = { https://doi.org/10.1007/s10460-023-10435-5},
author = {P Baur and A Iles},
}

@article{Martin2021,
title = {AV futures or futures with AVs? Bridging sociotechnical imaginaries and a multi-level perspective of autonomous vehicle visualisations in praxis},
journal = {Humanities and Social Sciences Communications},
volume = {8},
number = {68},
pages = {15},
year = {2021},
doi = {https://doi.org/10.1057/s41599-021-00739-4},
author = {Robert Martin},
}

@inproceedings{Mladenovic2020,
title = "Sociotechnical imaginaries of connected and automated vehicle technology: Comparative analysis of governance cultures in Finland, Germany, and the UK",
author = "Milos Mladenovic and Dominic Stead and Dimitris Milakis and Kate Pangbourne and Moshe Givoni",
year = "2020",
month = aug,
language = "English",
url = "https://bridgingtransport.org/",
booktitle = "Bridging Transportation Researchers Conference, BTR ; Conference date: 11-08-2020 Through 12-08-2020",
pages = {19},
publisher = {Virtual},
address = {Online},
}

@inproceedings{Eskenazi2023,
title = "Connected and Automated Vehicles as Tools for Sustainable Mobility Transitions? Sociotechnical Imaginaries of the French National CAV Strategy.",
author = "Manon Eskenazi",
year = "2023",
month = oct,
language = "English",
url = "https://hal.science/hal-04482098",
booktitle = "Global Mobility and Humanities Conference; Annual Conference of the International Association for the History of Transport, Traffic and Mobility",
pages = {},
publisher = {Academy of Mobility Humanities, Konkuk University},
address = {Seoul, South Korea},
}

@online{airside,
  author =       "Airport Consulting Partners",
organization =       "Airport Consulting Partners",
  title =        "Airside Facilities",
  url =          "https://www.airport-consult.com/en/center-of-excellence/business-areas/airside-facilities/#",
  lastaccessed = "Septermber 2, 2024",
year = "2024",
}

@inproceedings{Aljuneidi2024,
author = {Aljuneidi, Saja and Heuten, Wilko and Abdenebaoui, Larbi and Wolters, Maria K and Boll, Susanne},
title = {Why the Fine, AI? The Effect of Explanation Level on Citizens' Fairness Perception of AI-based Discretion in Public Administrations},
year = {2024},
isbn = {9798400703300},
publisher = {Association for Computing Machinery},
address = {New York, NY, USA},
url = {https://doi.org/10.1145/3613904.3642535},
doi = {10.1145/3613904.3642535},
booktitle = {Proceedings of the CHI Conference on Human Factors in Computing Systems},
articleno = {318},
numpages = {18},
location = {Honolulu, HI, USA},
series = {CHI '24}
}

@inproceedings{Freire2023,
author = {Kernan Freire, Samuel and Wang, Chaofan and Ruiz-Arenas, Santiago and Niforatos, Evangelos},
title = {Tacit Knowledge Elicitation for Shop-floor Workers with an Intelligent Assistant},
year = {2023},
isbn = {9781450394222},
publisher = {Association for Computing Machinery},
address = {New York, NY, USA},
url = {https://doi.org/10.1145/3544549.3585755},
doi = {10.1145/3544549.3585755},
booktitle = {Extended Abstracts of the 2023 CHI Conference on Human Factors in Computing Systems},
articleno = {266},
numpages = {7},
location = {Hamburg, Germany},
series = {CHI EA '23}
}

@article{Cabitza2020,
	author = {Federico Cabitza and Andrea Campagner and Clara Balsano},
	title = {Bridging the “last mile” gap between AI implementation and operation: “data awareness” that matters},
	journal = {Annals of Translational Medicine},
	volume = {8},
	number = {7},
	year = {2020},
numpages = {9},
	issn = {2305-5847},	url = {https://atm.amegroups.org/article/view/39228}
}

@inproceedings{Unhelkar2014,
author = {Unhelkar, Vaibhav V. and Siu, Ho Chit and Shah, Julie A.},
title = {Comparative performance of human and mobile robotic assistants in collaborative fetch-and-deliver tasks},
year = {2014},
isbn = {9781450326582},
publisher = {Association for Computing Machinery},
address = {New York, NY, USA},
url = {https://doi.org/10.1145/2559636.2559655},
doi = {10.1145/2559636.2559655},
booktitle = {Proceedings of the 2014 ACM/IEEE International Conference on Human-Robot Interaction},
pages = {82–89},
numpages = {8},
location = {Bielefeld, Germany},
series = {HRI '14}
}

@article{Russo2024,
author = {Russo, Daniel},
title = {Navigating the Complexity of Generative AI Adoption in Software Engineering},
year = {2024},
issue_date = {June 2024},
publisher = {Association for Computing Machinery},
address = {New York, NY, USA},
volume = {33},
number = {5},
issn = {1049-331X},
url = {https://doi.org/10.1145/3652154},
doi = {10.1145/3652154},
journal = {ACM Trans. Softw. Eng. Methodol.},
month = {jun},
articleno = {135},
numpages = {50}
}

@inproceedings{Iantorno2022,
author = {Iantorno, Mathew and Doggett, Olivia and Chandra, Priyank and Yujie Chen, Julie and Steup, Rosemary and Raval, Noopur and Khovanskaya, Vera and Lam, Laura and Singh, Anubha and Rotz, Sarah and Ratto, Matt},
title = {Outsourcing Artificial Intelligence: Responding to the Reassertion of the Human Element into Automation},
year = {2022},
isbn = {9781450391566},
publisher = {Association for Computing Machinery},
address = {New York, NY, USA},
url = {https://doi.org/10.1145/3491101.3503720},
doi = {10.1145/3491101.3503720},
booktitle = {Extended Abstracts of the 2022 CHI Conference on Human Factors in Computing Systems},
articleno = {103},
numpages = {5},
location = {New Orleans, LA, USA},
series = {CHI EA '22}
}

@inproceedings{Akridge2024,
author = {Akridge, Hunter and Fan, Bonnie and Tang, Alice Xiaodi and Mehta, Chinar and Martelaro, Nikolas and Fox, Sarah E},
title = {“The bus is nothing without us”: Making Visible the Labor of Bus Operators amid the Ongoing Push Towards Transit Automation},
year = {2024},
isbn = {9798400703300},
publisher = {Association for Computing Machinery},
address = {New York, NY, USA},
url = {https://doi.org/10.1145/3613904.3642714},
doi = {10.1145/3613904.3642714},
booktitle = {Proceedings of the CHI Conference on Human Factors in Computing Systems},
articleno = {479},
numpages = {16},
location = {Honolulu, HI, USA},
series = {CHI '24}
}

@article{Delfanti2021,
author = {Alessandro Delfanti and Bronwyn Frey},
title ={Humanly Extended Automation or the Future of Work Seen through Amazon Patents},
journal = {Science, Technology, \& Human Values},
volume = {46},
number = {3},
pages = {655-682},
year = {2021},
doi = {10.1177/0162243920943665},
URL = {       https://doi.org/10.1177/0162243920943665},
eprint = {   https://doi.org/10.1177/0162243920943665}

}

@inproceedings{Baxter2012,
author = {Baxter, Gordon and Rooksby, John and Wang, Yuanzhi and Khajeh-Hosseini, Ali},
title = {The ironies of automation: still going strong at 30?},
year = {2012},
isbn = {9781450317863},
publisher = {Association for Computing Machinery},
address = {New York, NY, USA},
url = {https://doi.org/10.1145/2448136.2448149},
doi = {10.1145/2448136.2448149},
booktitle = {Proceedings of the 30th European Conference on Cognitive Ergonomics},
pages = {65–71},
numpages = {7},
location = {Edinburgh, United Kingdom},
series = {ECCE '12}
}

@ARTICLE{Strauch2018,
  author={Strauch, Barry},
  journal={IEEE Transactions on Human-Machine Systems}, 
  title={Ironies of Automation: Still Unresolved After All These Years}, 
  year={2018},
  volume={48},
  number={5},
  pages={419-433},
  doi={10.1109/THMS.2017.2732506}}

@ARTICLE{Star1999,
  author={Star, S.L. and Strauss, A.},
  journal={Computer Supported Cooperative Work (CSCW)}, 
  title={Layers of Silence, Arenas of Voice: The Ecology of Visible and Invisible Work}, 
  year={1999},
  volume={8},
  number={},
  pages={9-30},
  doi={https://doi.org/10.1023/A:1008651105359}
}

@article{Suchman1995,
author = {Suchman, Lucy},
title = {Making work visible},
year = {1995},
issue_date = {Sept. 1995},
publisher = {Association for Computing Machinery},
address = {New York, NY, USA},
volume = {38},
number = {9},
issn = {0001-0782},
url = {https://doi.org/10.1145/223248.223263},
doi = {10.1145/223248.223263},
journal = {Commun. ACM},
month = {sep},
pages = {56–64},
numpages = {9}
}

@inproceedings{Gomez-Beldarrain2024,
author = {Gomez-Beldarrain, Garoa and Verma, Himanshu and Kim, Euiyoung and Bozzon, Alessandro},
title = {Revealing the Challenges to Automation Adoption in Organizations: Examining Practitioner Perspectives From an International Airport},
year = {2024},
isbn = {9798400703317},
publisher = {Association for Computing Machinery},
address = {New York, NY, USA},
url = {https://doi.org/10.1145/3613905.3650964},
doi = {10.1145/3613905.3650964},
booktitle = {Extended Abstracts of the 2024 CHI Conference on Human Factors in Computing Systems},
articleno = {281},
numpages = {7},
location = {
},
series = {CHI EA '24}
}

@article{Gamkrelidze2024,
author = {Gamkrelidze, T. and Zouinar, M. and Barcellini, F.},
title = {AI at work: understanding its uses and consequences on work activities and organization in radiology.},
year = {2024},
volume = {39},
number = {4},
doi = {https://doi.org/10.1007/s00146-024-01951-x},
journal = {AI and Society},
month = {may},
numpages = {19}
}

@book{Morozov2013,
    author = {Morozov, E},
    title = { To save everything, click here. The folly of technological solutionism.},
    publisher = {PublicAffairs},
address= {New York},
    year = 2013,
}

@inproceedings{Molin2024,
author = {Molin, Antonio},
title = {Examining Public Sector AI Adoption: Mechanisms for AI adoption in the absence of authoritative strategic direction},
year = {2024},
isbn = {9798400709883},
publisher = {Association for Computing Machinery},
address = {New York, NY, USA},
url = {https://doi.org/10.1145/3657054.3657278},
doi = {10.1145/3657054.3657278},
booktitle = {Proceedings of the 25th Annual International Conference on Digital Government Research},
pages = {764–775},
numpages = {12},
location = {Taipei, Taiwan},
series = {dg.o '24}
}

@inproceedings{Spektor2023DIS,
author = {Spektor, Franchesca and Fox, Sarah E and Awumey, Ezra and Riordan, Christine A. and Rho, Hye Jin and Kulkarni, Chinmay and Martinez-Lopez, Marlen and Stringam, Betsy and Begleiter, Ben and Forlizzi, Jodi},
title = {Designing for Wellbeing: Worker-Generated Ideas on Adapting Algorithmic Management in the Hospitality Industry},
year = {2023},
isbn = {9781450398930},
publisher = {Association for Computing Machinery},
address = {New York, NY, USA},
url = {https://doi.org/10.1145/3563657.3596018},
doi = {10.1145/3563657.3596018},
booktitle = {Proceedings of the 2023 ACM Designing Interactive Systems Conference},
pages = {623–637},
numpages = {15},
location = {Pittsburgh, PA, USA},
series = {DIS '23}
}

@Inbook{Fox2024_Humane,
author="Fox, Sarah E.
and Shorey, Samantha",
editor="Rousi, Rebekah
and von Koskull, Catharina
and Roto, Virpi",
title="Artfully Integrating AI: Proceeding Responsibly With Worker-Centered Best Practices",
bookTitle="Humane Autonomous Technology: Re-thinking Experience with and in Intelligent Systems",
year="2024",
publisher="Springer International Publishing",
address="Cham",
pages="67--84",
isbn="978-3-031-66528-8",
doi="10.1007/978-3-031-66528-8_4",
url="https://doi.org/10.1007/978-3-031-66528-8_4"
}

@book{Simonse2024,
  title={Design Roadmapping: Guidebook for Future Foresight Techniques},
  author={LWL Simonse},
  year={2024},
  publisher={TU Delft OPEN
Publishing},
address = {Delft, The Netherlands},
doi={10.59490/tb.84},
}

@article{Moradi2025,
author = {Moradi, Pegah and Levy, Karen and Cheyre, Cristobal},
title = {Pseudo-Automation: How Labor-Offsetting Technologies Reconfigure Roles and Relationships in Frontline Retail Work},
year = {2025},
issue_date = {May 2025},
publisher = {Association for Computing Machinery},
address = {New York, NY, USA},
volume = {9},
number = {2},
url = {https://doi.org/10.1145/3711051},
doi = {10.1145/3711051},
journal = {Proc. ACM Hum.-Comput. Interact.},
month = may,
articleno = {CSCW153},
numpages = {21}
}

@incollection{kuijer_automated_2019,
  title = {Automated {{Artefacts}} as {{Co-performers}} of {{Social Practices}}: {{Washing Machines}}, {{Laundering}} and {{Design}}},
  shorttitle = {Automated {{Artefacts}} as {{Co-performers}} of {{Social Practices}}},
  booktitle = {Social {{Practices}} and {{Dynamic Non-Humans}}: {{Nature}}, {{Materials}} and {{Technologies}}},
  author = {Kuijer, Lenneke},
  editor = {Maller, Cecily and Strengers, Yolande},
  year = {2019},
  pages = {193--214},
  publisher = {Springer International Publishing},
  address = {Cham},
  doi = {10.1007/978-3-319-92189-1_10},
  urldate = {2025-06-30},
  isbn = {978-3-319-92189-1},
  langid = {english},
}

@inproceedings{kuijer_co-performance_2018,
  title = {Co-Performance: {{Conceptualizing}} the {{Role}} of {{Artificial Agency}} in the {{Design}} of {{Everyday Life}}},
  shorttitle = {Co-Performance},
  booktitle = {Proceedings of the 2018 {{CHI Conference}} on {{Human Factors}} in {{Computing Systems}}},
  author = {Kuijer, Lenneke and Giaccardi, Elisa},
  year = {2018},
  month = apr,
  series = {{{CHI}} '18},
  pages = {1--13},
  publisher = {Association for Computing Machinery},
  address = {New York, NY, USA},
  doi = {10.1145/3173574.3173699},
  urldate = {2025-06-30},
  isbn = {978-1-4503-5620-6},
}

@article{van_beek_everyday_2025,
  title = {The Everyday Enactment of Interfaces: A Study of Crises and Conflicts in the More-than-Human Home},
  shorttitle = {The Everyday Enactment of Interfaces},
  author = {{van Beek}, Evert and , Elisa, Giaccardi and , Stella, Boess and {and Bozzon}, Alessandro},
  year = {2025},
  month = jan,
  journal = {Human--Computer Interaction},
  volume = {40},
  number = {1-4},
  pages = {221--248},
  publisher = {Taylor \& Francis},
  issn = {0737-0024},
  doi = {10.1080/07370024.2023.2283536},
  urldate = {2025-06-30},
}

@inproceedings{van_beek_making_2023,
  title = {Making a Scene: {{Representing}} and Annotating Enacted Interfaces in Co-Performances Using the Screenplay},
  shorttitle = {Making a Scene},
  booktitle = {{{IASDR}} 2023: {{Life-Changing Design}}},
  author = {van Beek, Evert and Giaccardi, Elisa and Boess, Stella and Bozzon, Alessandro},
  editor = {De Sainz Molestina, Davide and Galluzzo, Laura and Rizzo, Francesca and Spallazzo, Davide},
  year = {2023},
  month = oct,
  pages = {12},
  address = {Milan, Italy},
  publisher = {Design Research Society},
  doi = {10.21606/iasdr.2023.788},
  url = {https://doi.org/10.21606/iasdr.2023.788},
  urldate = {2025-07-19}
}

@inproceedings{GomezBeldarrain2025,
author = {Gomez-Beldarrain, Garoa and Verma, Himanshu and Kim, Euiyoung and Bozzon, Alessandro},
title = {Why does Automation Adoption in Organizations Remain a Fallacy?: Scrutinizing Practitioners' Imaginaries in an International Airport},
year = {2025},
isbn = {9798400713941},
publisher = {Association for Computing Machinery},
address = {New York, NY, USA},
url = {https://doi.org/10.1145/3706598.3713978},
doi = {10.1145/3706598.3713978},
booktitle = {Proceedings of the 2025 CHI Conference on Human Factors in Computing Systems},
articleno = {213},
numpages = {19},
location = {
},
series = {CHI '25}
}

@inproceedings{Riek2025,
author = {Riek, Laurel D. and Irani, Lilly},
title = {The Future Is Rosie?: Disempowering Arguments About Automation and What to Do About It},
year = {2025},
isbn = {9798400713941},
publisher = {Association for Computing Machinery},
address = {New York, NY, USA},
url = {https://doi.org/10.1145/3706598.3714151},
doi = {10.1145/3706598.3714151},
booktitle = {Proceedings of the 2025 CHI Conference on Human Factors in Computing Systems},
articleno = {575},
numpages = {14},
location = {
},
series = {CHI '25}
}

@article{Humphreys2005,
   author = {Lee Humphreys},
   doi = {10.1080/02691720500145449},
   issn = {02691728},
   issue = {2-3},
   journal = {Social Epistemology},
   pages = {231-253},
   publisher = {Routledge},
   title = {Reframing social groups, closure, and stabilization in the social construction of technology},
   volume = {19},
   year = {2005},
}

@article{Pinch1984,
   author = {Trevor J. Pinch and Wiebe E. Bijker},
   doi = {10.1177/030631284014003004},
   issn = {14603659},
   issue = {3},
   journal = {Social Studies of Science},
   pages = {399-441},
   title = {The Social Construction of Facts and Artefacts: Or How the Sociology of Science and the Sociology of Technology might Benefit Each Other},
   volume = {14},
   year = {1984},
}

@article{braun_can_2021,
  title = {Can {{I}} Use {{TA}}? {{Should I}} Use {{TA}}? {{Should I}} {\emph{Not}} Use {{TA}}? {{Comparing}} Reflexive Thematic Analysis and Other Pattern-based Qualitative Analytic Approaches},
  shorttitle = {Can {{I}} Use {{TA}}?},
  author = {Braun, Virginia and Clarke, Victoria},
  year = {2021},
  month = mar,
  journal = {Counselling and Psychotherapy Research},
  volume = {21},
  number = {1},
  pages = {37--47},
  issn = {1473-3145, 1746-1405},
  doi = {10/ghf388},
  urldate = {2022-12-12},
  langid = {english}
}

@article{braun_conceptual_2022,
  title = {Conceptual and Design Thinking for Thematic Analysis.},
  author = {Braun, Virginia and Clarke, Victoria},
  year = {2022},
  month = feb,
  journal = {Qualitative Psychology},
  volume = {9},
  number = {1},
  pages = {3--26},
  issn = {2326-3598, 2326-3601},
  doi = {10/gj2m7c},
  urldate = {2023-06-07},
  langid = {english},
}

@article{braun_reflecting_2019,
  title = {Reflecting on Reflexive Thematic Analysis},
  author = {Braun, Virginia and Clarke, Victoria},
  year = {2019},
  month = aug,
  journal = {Qualitative Research in Sport, Exercise and Health},
  volume = {11},
  number = {4},
  pages = {589--597},
  issn = {2159-676X, 2159-6778},
  doi = {10/gf89jz},
  urldate = {2023-03-07},
  langid = {english},
}

@article{braun_using_2006,
  title = {Using Thematic Analysis in Psychology},
  author = {Braun, Virginia and Clarke, Victoria},
  year = {2006},
  journal = {Qualitative Research in Psychology},
  volume = {3},
  number = {2},
  pages = {77--101},
  issn = {14780887},
  doi = {10/fswdcx},
}

@incollection{cooper_thematic_2012,
  title = {Thematic Analysis.},
  booktitle = {{{APA}} Handbook of Research Methods in Psychology, {{Vol}} 2: {{Research}} Designs: {{Quantitative}}, Qualitative, Neuropsychological, and Biological.},
  author = {Braun, Virginia and Clarke, Victoria},
  editor = {Cooper, Harris and Camic, Paul M. and Long, Debra L. and Panter, A. T. and Rindskopf, David and Sher, Kenneth J.},
  year = {2012},
  pages = {57--71},
  publisher = {American Psychological Association},
  address = {Washington},
  doi = {10.1037/13620-004},
  urldate = {2022-06-30},
  isbn = {978-1-4338-1005-3},
  langid = {english},
}

@Inbook{Roto2024,
author="Roto, Virpi",
editor="Rousi, Rebekah
and von Koskull, Catharina
and Roto, Virpi",
title="Co-worker, Butler, or Coach? Designing Automation for Work Enrichment",
bookTitle="Humane Autonomous Technology: Re-thinking Experience with and in Intelligent Systems",
year="2024",
publisher="Springer International Publishing",
address="Cham",
pages="45--65",
isbn="978-3-031-66528-8",
doi="10.1007/978-3-031-66528-8_3",
url="https://doi.org/10.1007/978-3-031-66528-8_3"
}

@article{Johnsson2017_highperf,
author = {Johnsson, Mikael},
year = {2017},
month = {12},
pages = {},
title = {Creating High-performing Innovation Teams},
volume = {5},
journal = {Journal of Innovation Management},
doi = {10.24840/2183-0606_005.004_0004}
}

@book{Gode2020,
author = {Both, Göde},
year = {2020},
month = {04},
pages = {},
title = {Keeping Autonomous Driving Alive: An Ethnography of Visions, Masculinity and Fragility},
isbn = {9783966650090},
doi = {10.3224/96665009}
}

@article{Morosan2022,
    author = {Morosan, Cristian and Bowen, John T.},
    title = {Labor shortage solution: redefining hospitality through digitization},
    journal = {International Journal of Contemporary Hospitality Management},
    volume = {34},
    number = {12},
    pages = {4674-4685},
    year = {2022},
    month = {06},
    issn = {0959-6119},
    doi = {10.1108/IJCHM-03-2022-0304},
    url = {https://doi.org/10.1108/IJCHM-03-2022-0304},
    eprint = {https://www.emerald.com/ijchm/article-pdf/34/12/4674/1093516/ijchm-03-2022-0304.pdf},
}

@article{Maibaum2022,
    author = {Maibaum, A. and Bischof, A. and Hergesell, J. and Lipp, B.},
    title = { A critique of robotics in health care.},
    journal = {AI and Society},
    volume = {37},
    pages = {467–477},
    year = {2022},
    month = {04},
    url = {https://doi.org/10.1007/s00146-021-01206-z},
}

@article{Valles-Peris2020,
author = {Núria Vallès-Peris and Miquel Domènech},
title = {Roboticists’ Imaginaries of Robots for Care: The Radical Imaginary as a Tool for an Ethical Discussion},
journal = {Engineering Studies},
volume = {12},
number = {3},
pages = {157--176},
year = {2020},
publisher = {Routledge},
doi = {10.1080/19378629.2020.1821695},
URL = { https://doi.org/10.1080/19378629.2020.1821695
},
eprint = { https://doi.org/10.1080/19378629.2020.1821695}
}

@article{Leonardi2011,
   author = {Paul M Leonardi},
   issue = {1},
   journal = {Source: MIS Quarterly},
   pages = {147-167},
   title = {When Flexible Routines Meet Flexible Technologies: Affordance, Constraint, and the Imbrication of Human and Material Agencies},
   volume = {35},
   year = {2011},
}

@article{Orlikowski2007,
   author = {Wanda J. Orlikowski},
   doi = {10.1177/0170840607081138},
   issn = {01708406},
   issue = {9},
   journal = {Organization Studies},
   month = {9},
   pages = {1435-1448},
   title = {Sociomaterial practices: Exploring technology at work},
   volume = {28},
   year = {2007},
}

@article{Orlikowski2008,
   author = {Wanda J. Orlikowski and Susan V. Scott},
   doi = {10.1080/19416520802211644},
   issn = {1941-6520},
   issue = {1},
   journal = {The Academy of Management Annals},
   month = {1},
   pages = {433-474},
   publisher = {Academy of Management},
   title = {10 Sociomateriality: Challenging the Separation of Technology, Work and Organization},
   volume = {2},
   year = {2008},
}

@article{Bailey2022,
   author = {Diane E. Bailey and Samer Faraj and Pamela J. Hinds and Paul M. Leonardi and Georg von Krogh},
   doi = {10.1287/ORSC.2021.1562},
   issn = {15265455},
   issue = {1},
   journal = {Organization Science},
   month = {1},
   pages = {1-18},
   publisher = {INFORMS Inst.for Operations Res.and the Management Sciences},
   title = {We Are All Theorists of Technology Now: A Relational Perspective on Emerging Technology and Organizing},
   volume = {33},
   year = {2022},
}

@article{Leonardi2013,
   author = {Paul M. Leonardi},
   doi = {10.1016/j.infoandorg.2013.02.002},
   issn = {14717727},
   issue = {2},
   journal = {Information and Organization},
   month = {4},
   pages = {59-76},
   title = {Theoretical foundations for the study of sociomateriality},
   volume = {23},
   year = {2013},
}

@article{Leonardi2008,
   author = {Paul M. Leonardi and Stephen R. Barley},
   doi = {10.1016/j.infoandorg.2008.03.001},
   issn = {14717727},
   issue = {3},
   journal = {Information and Organization},
   pages = {159-176},
   title = {Materiality and change: Challenges to building better theory about technology and organizing},
   volume = {18},
   year = {2008},
}

@article{Orlikowski1992,
   author = {Wanda J Orlikowski},
   issue = {3},
   journal = {Source: Organization Science},
   pages = {398-427},
   title = {The Duality of Technology: Rethinking the Concept of Technology in Organizations},
   volume = {3},
   url = {https://about.jstor.org/terms},
   year = {1992},
}

@inbook{Jackson2002,
   author = {Jackson, M. and Poole, M. and Kuhn, T.},
   doi = {https://doi.org/10.4135/9781848608245.n18},
   bookTitle = {The social construction of technology in studies of the workplace, L. Lievrouw and S. Livingstone (Eds.)},
 
   pages = {236-253},
   title = {The social construction of technology in studies of the workplace},
 publisher= {SAGE Publications, Ltd},
   year = {2002},
}

@article{Klitsie2019,
author = {Klitsie, Joannes Barend and Price, Rebecca Anne and De Lille, Christine Stefanie Heleen},
title = {Overcoming the Valley of Death: A Design Innovation Perspective},
journal = {Design Management Journal},
volume = {14},
number = {1},
pages = {28-41},
doi = {https://doi.org/10.1111/dmj.12052},
url = {https://onlinelibrary.wiley.com/doi/abs/10.1111/dmj.12052},
eprint = {https://onlinelibrary.wiley.com/doi/pdf/10.1111/dmj.12052},
year = {2019}
}

@inproceedings{Shukla2025,
author = {Shukla, Prakash and Bui, Phuong and Levy, Sean S and Kowalski, Max and Baigelenov, Ali and Parsons, Paul},
title = {De-skilling, Cognitive Offloading, and Misplaced Responsibilities: Potential Ironies of AI-Assisted Design},
year = {2025},
isbn = {9798400713958},
publisher = {Association for Computing Machinery},
address = {New York, NY, USA},
url = {https://doi.org/10.1145/3706599.3719931},
doi = {10.1145/3706599.3719931},
booktitle = {Proceedings of the Extended Abstracts of the CHI Conference on Human Factors in Computing Systems},
articleno = {171},
numpages = {7},
location = {
},
series = {CHI EA '25}
}

@article{MontielValle2024,
author = {Dominique A. Montiel Valle and Samantha Shorey},
title = {Dis//assemblages of AI: repair labor and resistance in the automated workplace},
journal = {Information, Communication \& Society},
volume = {27},
number = {10},
pages = {2022--2037},
year = {2024},
publisher = {Routledge},
doi = {10.1080/1369118X.2024.2371794},
URL = { https://doi.org/10.1080/1369118X.2024.2371794},
eprint = { https://doi.org/10.1080/1369118X.2024.2371794},

}

@inproceedings{Ma2025,
author = {Ma, Shuhao and Liu, Zhiming and Nisi, Valentina and Fox, Sarah E and Nunes, Nuno Jardim},
title = {Speculative Job Design: Probing Alternative Opportunities for Gig Workers in an Automated Future},
year = {2025},
isbn = {9798400713941},
publisher = {Association for Computing Machinery},
address = {New York, NY, USA},
url = {https://doi.org/10.1145/3706598.3713885},
doi = {10.1145/3706598.3713885},
booktitle = {Proceedings of the 2025 CHI Conference on Human Factors in Computing Systems},
articleno = {1208},
numpages = {18},
location = {
},
series = {CHI '25}
}

@article{Perry2005,
author = {Perry, Shawna J. MD and Wears, Robert L. MD, MS and Cook, Richard I. MD.},
title = {The Role of Automation in Complex System Failures},
journal = {Journal of Patient Safety},
volume = {1},
number = {1},
pages = {56-61},
year = {2005},

doi = {10.1097/01209203-200503000-00010},
}

@article{Demirci2021,
author = {Demirci, Seref},
year = {2021},
month = {08},
pages = {},
title = {The requirements for automation systems based on Boeing 737 MAX crashes},
volume = {ahead-of-print},
journal = {Aircraft Engineering and Aerospace Technology},
doi = {10.1108/AEAT-03-2021-0069}
}

@ARTICLE{Bradshaw2014_2,
  author={Hoffman, Robert R. and Hawley, John K. and Bradshaw, Jeffrey M.},
  journal={IEEE Intelligent Systems}, 
  title={Myths of Automation, Part 2: Some Very Human Consequences}, 
  year={2014},
  volume={29},
  number={2},
  pages={82-85},
  doi={10.1109/MIS.2014.25}}

@article{Acemoglu2019,
Author = {Acemoglu, Daron and Restrepo, Pascual},
Title = {Automation and New Tasks: How Technology Displaces and Reinstates Labor},
Journal = {Journal of Economic Perspectives},
Volume = {33},
Number = {2},
Year = {2019},
Month = {May},
Pages = {3–30},
DOI = {10.1257/jep.33.2.3},
URL = {https://www.aeaweb.org/articles?id=10.1257/jep.33.2.3}}

@article{hellstrom2002,
    author = {Hellström, Tomas and Jacob, Merle and Malmquist, Ulf},
    title = {Guiding innovation socially and cognitively: the innovation team model at Skanova Networks},
    journal = {European Journal of Innovation Management},
    volume = {5},
    number = {3},
    pages = {172-180},
    year = {2002},
    month = {09},
    issn = {1460-1060},
    doi = {10.1108/14601060210436745},
    url = {https://doi.org/10.1108/14601060210436745},
    eprint = {https://www.emerald.com/ejim/article-pdf/5/3/172/267827/14601060210436745.pdf},
}

@article{DAMANPOUR2006_inn,
title = {Research on innovation in organizations: Distinguishing innovation-generating from innovation-adopting organizations},
journal = {Journal of Engineering and Technology Management},
volume = {23},
number = {4},
pages = {269-291},
year = {2006},
issn = {0923-4748},
doi = {https://doi.org/10.1016/j.jengtecman.2006.08.002},
url = {https://www.sciencedirect.com/science/article/pii/S0923474806000403},
author = {Fariborz Damanpour and J. Daniel Wischnevsky}
}

@book{dourish_where_2004,
  title = {Where the Action Is: The Foundations of Embodied Interaction},
  shorttitle = {Where the Action Is},
  author = {Dourish, Paul},
  year = {2004},
  series = {A {{Bradford}} Book},
  edition = {1. MIT Press paperback ed},
  publisher = {MIT Press},
  address = {Cambridge, Mass. London},
  isbn = {978-0-262-54178-7 978-0-262-04196-6},
  langid = {english}
}

@book{suchman_human-machines_2009,
  title = {Human-Machines Reconfigurations: Plans and Situated Actions},
  shorttitle = {Human-Machines Reconfigurations},
  author = {Suchman, Lucy},
  year = {2009},
  edition = {2nd ed},
  publisher = {Cambridge University Press},
  address = {New York},
  isbn = {978-0-521-67588-8},
  langid = {english},
  lccn = {004.019}
}

@inproceedings{Ma2023,
author = {Ma, Shuhao and Bala, Paulo and Nisi, Valentina and Zimmerman, John and Nunes, Nuno Jardim},
title = {Uncovering Gig Worker-Centered Design Opportunities in Food Delivery Work},
year = {2023},
isbn = {9781450398930},
publisher = {Association for Computing Machinery},
address = {New York, NY, USA},
url = {https://doi.org/10.1145/3563657.3596123},
doi = {10.1145/3563657.3596123},
booktitle = {Proceedings of the 2023 ACM Designing Interactive Systems Conference},
pages = {688–701},
numpages = {14},
location = {Pittsburgh, PA, USA},
series = {DIS '23}
}

@article{Bassett2019,
author = {Bassett, Caroline and Roberts, Ben},
year = {2019},
month = {07},
pages = {9-28},
title = {Automation Now and Then: Automation Fevers, Anxieties and Utopias},
volume = {98},
journal = {New Formations},
doi = {10.3898/NEWF:98.02.2019}
}

@inproceedings{George1990,
author = {George, Joey F. and Valacich, Joseph S. and Nunamaker, J. F.},
title = {The organizational implementation of an electronic meeting system: an analysis of the innovation process},
year = {1990},
isbn = {0201509326},
publisher = {Association for Computing Machinery},
address = {New York, NY, USA},
url = {https://doi.org/10.1145/97243.97308},
doi = {10.1145/97243.97308},
booktitle = {Proceedings of the SIGCHI Conference on Human Factors in Computing Systems},
pages = {361–368},
numpages = {8},
location = {Seattle, Washington, USA},
series = {CHI '90}
}

@inproceedings{Yildirim2024,
author = {Yildirim, Nur and Zlotnikov, Susanna and Sayar, Deniz and Kahn, Jeremy M. and Bukowski, Leigh A and Amin, Sher Shah and Riman, Kathryn A. and Davis, Billie S. and Minturn, John S. and King, Andrew J. and Ricketts, Dan and Tang, Lu and Sivaraman, Venkatesh and Perer, Adam and Preum, Sarah M. and McCann, James and Zimmerman, John},
title = {Sketching AI Concepts with Capabilities and Examples: AI Innovation in the Intensive Care Unit},
year = {2024},
isbn = {9798400703300},
publisher = {Association for Computing Machinery},
address = {New York, NY, USA},
url = {https://doi.org/10.1145/3613904.3641896},
doi = {10.1145/3613904.3641896},
booktitle = {Proceedings of the 2024 CHI Conference on Human Factors in Computing Systems},
articleno = {451},
numpages = {18},
location = {Honolulu, HI, USA},
series = {CHI '24}
}

@inproceedings{Waddell2024,
author = {Waddell, Alex and Seguin, Joshua Paolo and Wu, Ling and Stragalinos, Peta and Wherton, Joe and Watterson, Jessica L and Prawira, Christopher Owen and Olivier, Patrick and Manning, Victoria and Lubman, Dan and Grigg, Jasmin},
title = {Leveraging Implementation Science in Human-Centred Design for Digital Health},
year = {2024},
isbn = {9798400703300},
publisher = {Association for Computing Machinery},
address = {New York, NY, USA},
url = {https://doi.org/10.1145/3613904.3642161},
doi = {10.1145/3613904.3642161},
booktitle = {Proceedings of the 2024 CHI Conference on Human Factors in Computing Systems},
articleno = {240},
numpages = {17},
location = {Honolulu, HI, USA},
series = {CHI '24}
}

@inproceedings{Zajkac2024,
author = {Zaj\k{a}c, Hubert Dariusz and Ribeiro, Jorge Miguel Neves and Ingala, Silvia and Gentile, Simona and Wanjohi, Ruth and Gitau, Samuel Nguku and Carlsen, Jonathan Frederik and Nielsen, Michael Bachmann and Andersen, Tariq Osman},
title = {"It depends": Configuring AI to Improve Clinical Usefulness Across Contexts},
year = {2024},
isbn = {9798400705830},
publisher = {Association for Computing Machinery},
address = {New York, NY, USA},
url = {https://doi.org/10.1145/3643834.3660707},
doi = {10.1145/3643834.3660707},
booktitle = {Proceedings of the 2024 ACM Designing Interactive Systems Conference},
pages = {874–889},
numpages = {16},
location = {Copenhagen, Denmark},
series = {DIS '24}
}

@article{Zajac2023,
author = {Zaj\k{a}c, Hubert D. and Li, Dana and Dai, Xiang and Carlsen, Jonathan F. and Kensing, Finn and Andersen, Tariq O.},
title = {Clinician-Facing AI in the Wild: Taking Stock of the Sociotechnical Challenges and Opportunities for HCI},
year = {2023},
issue_date = {April 2023},
publisher = {Association for Computing Machinery},
address = {New York, NY, USA},
volume = {30},
number = {2},
issn = {1073-0516},
url = {https://doi.org/10.1145/3582430},
doi = {10.1145/3582430},
journal = {ACM Trans. Comput.-Hum. Interact.},
month = mar,
articleno = {33},
numpages = {39}
}

@article{Zajac2025,
author = {Zaj\k{a}c, Hubert D. and Andersen, Tariq O. and Kwasa, Elijah and Wanjohi, Ruth and Onyinkwa, Mary K. and Mwaniki, Edward K. and Gitau, Samuel N. and Yaseen, Shawnim S. and Carlsen, Jonathan F. and Fraccaro, Marco and Nielsen, Michael B. and Chen, Yunan},
title = {Towards Clinically Useful AI: From Radiology Practices in Global South and North to Visions of AI Support},
year = {2025},
issue_date = {April 2025},
publisher = {Association for Computing Machinery},
address = {New York, NY, USA},
volume = {32},
number = {2},
issn = {1073-0516},
url = {https://doi.org/10.1145/3715115},
doi = {10.1145/3715115},
journal = {ACM Trans. Comput.-Hum. Interact.},
month = apr,
articleno = {20},
numpages = {38}
}

@inproceedings{alfrink_contestable_2023-2,
  title = {Contestable {{Camera Cars}}: {{A Speculative Design Exploration}} of {{Public AI That Is Open}} and {{Responsive}} to {{Dispute}}},
  shorttitle = {Contestable {{Camera Cars}}},
  booktitle = {Proceedings of the 2023 {{CHI Conference}} on {{Human Factors}} in {{Computing Systems}}},
  author = {Alfrink, Kars and Keller, Ianus and Doorn, Neelke and Kortuem, Gerd},
  year = {2023},
  month = apr,
  series = {{{CHI}} '23},
  pages = {1--16},
  publisher = {Association for Computing Machinery},
  address = {New York, NY, USA},
  doi = {10/gr5wcx},
  urldate = {2025-09-11},
  isbn = {978-1-4503-9421-5},
}

@inproceedings{claisse_keeping_2023,
  title = {`{{Keeping}} Our {{Faith Alive}}': {{Investigating Buddhism Practice}} during {{COVID-19}} to {{Inform Design}} for the {{Online Community Practice}} of {{Faith}}},
  shorttitle = {`{{Keeping}} Our {{Faith Alive}}'},
  booktitle = {Proceedings of the 2023 {{CHI Conference}} on {{Human Factors}} in {{Computing Systems}}},
  author = {Claisse, Caroline and Durrant, Abigail C},
  year = {2023},
  month = apr,
  series = {{{CHI}} '23},
  pages = {1--19},
  publisher = {Association for Computing Machinery},
  address = {New York, NY, USA},
  doi = {10/g82ft7},
  urldate = {2025-09-11},
  isbn = {978-1-4503-9421-5},
}

@inproceedings{offerman_rediscovering_2025,
  title = {({{Re}})Discovering {{Sexual Pleasure}} after {{Cancer}}: {{Understanding}} the {{Design Space}}},
  shorttitle = {({{Re}})Discovering {{Sexual Pleasure}} after {{Cancer}}},
  booktitle = {Proceedings of the 2025 {{CHI Conference}} on {{Human Factors}} in {{Computing Systems}}},
  author = {Offerman, C{\'e}line and Bourgeois, Jacky and {van Beurden}, Jules and Bozzon, Alessandro},
  year = {2025},
  month = apr,
  series = {{{CHI}} '25},
  pages = {1--17},
  publisher = {Association for Computing Machinery},
  address = {New York, NY, USA},
  doi = {10/g93bx5},
  urldate = {2025-09-11},
  isbn = {979-8-4007-1394-1},
}

@InProceedings{Kallbacker2025,
author="K{\"a}llb{\"a}cker, Jonathan
and Bergqvist, Andreas
and Cajander, {\AA}sa
and Cort, Rebecca
and Normark, Maria
and Premanandan, Shweta",
editor="Streitz, Norbert A.
and Konomi, Shin'ichi",
title="Exploring the Future of AI, Autonomous Vehicles, and Emerging Technologies in Airport Operations: Stakeholder Perspectives",
booktitle="Distributed, Ambient and Pervasive Interactions",
year="2025",
publisher="Springer Nature Switzerland",
address="Cham",
pages="3--20",
isbn="978-3-031-92980-9"
}

@article{Turtle2025,
  title = {Undoing Gracia: queering the self in the algorithmic borderlands},
  author = {Turtle, G.L. and Giaccardi, E. and Bendor, R.},
  year = 2025,
  journal = {AI and Society},
  pages = {18},
  doi = {https://doi.org/10.1007/s00146-025-02661-8},
}

@article{Bomba2024,
author = {Bomba, Federico and Men\'{e}ndez-Blanco, Mar\'{\i}a and Grigis, Paolo and Cremaschi, Michele and De Angeli, Antonella},
title = {The Choreographer-Performer Continuum: A Diffraction Tool to Illuminate Authorship in More Than Human Co-Performances},
year = {2024},
issue_date = {December 2024},
publisher = {Association for Computing Machinery},
address = {New York, NY, USA},
volume = {31},
number = {6},
issn = {1073-0516},
url = {https://doi.org/10.1145/3689040},
doi = {10.1145/3689040},
journal = {ACM Trans. Comput.-Hum. Interact.},
month = dec,
articleno = {75},
numpages = {23}
}

@inproceedings{KimLim2019,
author = {Kim, Da-jung and Lim, Youn-kyung},
title = {Co-Performing Agent: Design for Building User-Agent Partnership in Learning and Adaptive Services},
year = {2019},
isbn = {9781450359702},
publisher = {Association for Computing Machinery},
address = {New York, NY, USA},
url = {https://doi.org/10.1145/3290605.3300714},
doi = {10.1145/3290605.3300714},
booktitle = {Proceedings of the 2019 CHI Conference on Human Factors in Computing Systems},
pages = {1–14},
numpages = {14},
location = {Glasgow, Scotland Uk},
series = {CHI '19}
}

@inproceedings{Kuijer2022,
author = {Kuijer, L. and Hensen Centnerová, L.},
title = {Exploring futures of summer comfort in Dutch households.},
year = {2022},
doi = {https://doi.org/10.34641/clima.2022.388},
booktitle = {Proceedings of the CLIMA 2022 conference},
numpages = {8},
}

@inproceedings{Nicenboim2022,
author = {Nicenboim, I. and Giaccardi, E. and Redström, J.},
title = {From explanations to shared understandings of AI},
year = {2022},
doi = {https://doi.org/10.21606/drs.2022.773},
booktitle = { in Lockton, D., Lenzi, S., Hekkert, P., Oak, A., Sádaba, J., Lloyd, P. (eds.), DRS2022: Bilbao, 25 June - 3 July, Bilbao, Spain. },
numpages = {14},
}

@ARTICLE{Viaene2021,
AUTHOR={Viaene, Emilia  and Kuijer, Lenneke  and Funk, Mathias },    
TITLE={Learning Systems versus Future Everyday Domestic Life: A Designer’s Interpretation of Social Practice Imaginaries},       
JOURNAL={Frontiers in Artificial Intelligence},        
VOLUME={Volume 4 - 2021},
YEAR={2021},
URL={https://www.frontiersin.org/journals/artificial-intelligence/articles/10.3389/frai.2021.707562},
DOI={10.3389/frai.2021.707562},
ISSN={2624-8212},}

@book{Schatzki2002,
    author = "Schatzki, Theodore",
    title = "The Site of the Social: A Philosophical Account of the Constitution of Social Life and Change",
    publisher = "Penn State University Press",
    year = 2002 
}

@book{Shove2012,
    author = "Shove, Elizabeth and Mika Pantzar and Matt Watson",
    title = "The Dynamics of Social
Practice: Everyday Life and How It Changes",
    publisher = "London: Sage",
    year = "2012"
}

@article{Reckwitz2002,
    author = "Reckwitz, A.",
    title = "Toward a Theory of Social Practices: A Development in Culturalist Theorizing" ,
    journal = "European Journal of Social Theory",
volume = "5",
number = "2",
pages = "243-263",
    year = "2002",
doi = "https://doi.org/10.1177/13684310222225432"
}

@article{giaccardi_technology_2020,
  title = {Technology and {{More-Than-Human Design}}},
  author = {Giaccardi, Elisa and Redstr{\"o}m, Johan},
  year = 2020,
  month = sep,
  journal = {Design Issues},
  volume = {36},
  number = {4},
  pages = {33--44},
  issn = {0747-9360, 1531-4790},
  doi = {10/gh6rvm},
  urldate = {2022-06-03},
  langid = {english}
}

@inproceedings{Garret2025,
author = {Garrett, Rachael and Brundell, Patrick and Castle-Green, Simon and Hawkins, Kat and Tennent, Paul and Zhou, Feng and Lampinen, Airi and H\"{o}\"{o}k, Kristina and Benford, Steve},
title = {Friction in Processual Ethics: Reconfiguring Ethical Relations in Interdisciplinary Research},
year = {2025},
isbn = {9798400713941},
publisher = {Association for Computing Machinery},
address = {New York, NY, USA},
url = {https://doi.org/10.1145/3706598.3714123},
doi = {10.1145/3706598.3714123},
booktitle = {Proceedings of the 2025 CHI Conference on Human Factors in Computing Systems},
articleno = {400},
numpages = {15},
location = {
},
series = {CHI '25}
}

@inproceedings{Gould2021,
author = {Gould, Sandy J. J. and Chuang, Lewis L. and Iacovides, Ioanna and Garaialde, Diego and Cecchinato, Marta E. and Cowan, Benjamin R. and Cox, Anna L.},
title = {A Special Interest Group on Designed and Engineered Friction in Interaction},
year = {2021},
isbn = {9781450380959},
publisher = {Association for Computing Machinery},
address = {New York, NY, USA},
url = {https://doi.org/10.1145/3411763.3450404},
doi = {10.1145/3411763.3450404},
booktitle = {Extended Abstracts of the 2021 CHI Conference on Human Factors in Computing Systems},
articleno = {158},
numpages = {4},
location = {Yokohama, Japan},
series = {CHI EA '21}
}

@inproceedings{Haliburton2024,
author = {Haliburton, Luke and Gr\"{u}ning, David Joachim and Riedel, Frederik and Schmidt, Albrecht and Terzimehi\'{c}, Na\dj{}a},
title = {A Longitudinal In-the-Wild Investigation of Design Frictions to Prevent Smartphone Overuse},
year = {2024},
isbn = {9798400703300},
publisher = {Association for Computing Machinery},
address = {New York, NY, USA},
url = {https://doi.org/10.1145/3613904.3642370},
doi = {10.1145/3613904.3642370},
booktitle = {Proceedings of the 2024 CHI Conference on Human Factors in Computing Systems},
articleno = {243},
numpages = {16},
location = {Honolulu, HI, USA},
series = {CHI '24}
}

@inproceedings{Sanchez2025,
author = {Sanchez, Camilo and Wang, Sui and Savolainen, Kaisa and Epp, Felix Anand and Salovaara, Antti},
title = {Let's Talk Futures: A Literature Review of HCI's Future Orientation},
year = {2025},
isbn = {9798400713941},
publisher = {Association for Computing Machinery},
address = {New York, NY, USA},
url = {https://doi.org/10.1145/3706598.3713759},
doi = {10.1145/3706598.3713759},
booktitle = {Proceedings of the 2025 CHI Conference on Human Factors in Computing Systems},
articleno = {487},
numpages = {36},
location = {
},
series = {CHI '25}
}

@inproceedings{Salovaara2025,
author = {Salovaara, Antti and Vahvelainen, Leevi},
title = {Triangulating on Possible Futures: Conducting User Studies on Several Futures Instead of Only One},
year = {2025},
isbn = {9798400713941},
publisher = {Association for Computing Machinery},
address = {New York, NY, USA},
url = {https://doi.org/10.1145/3706598.3713565},
doi = {10.1145/3706598.3713565},
booktitle = {Proceedings of the 2025 CHI Conference on Human Factors in Computing Systems},
articleno = {478},
numpages = {16},
location = {
},
series = {CHI '25}
}

@inproceedings{Salovaara2022,
author = {Moesgen, Tim and Epp, Felix Anand and Salovaara, Antti and Pouta, Emmi and Sanchez, Camilo},
title = {Hands-on Introduction to Futures Thinking and Foresight with the Future Ripples Method},
year = {2022},
isbn = {9781450394482},
publisher = {Association for Computing Machinery},
address = {New York, NY, USA},
url = {https://doi.org/10.1145/3547522.3558902},
doi = {10.1145/3547522.3558902},
booktitle = {Adjunct Proceedings of the 2022 Nordic Human-Computer Interaction Conference},
articleno = {20},
numpages = {2},
location = {Aarhus, Denmark},
series = {NordiCHI '22 Adjunct}
}

@inproceedings{Epp2022,
author = {Epp, Felix Anand and Moesgen, Tim and Salovaara, Antti and Pouta, Emmi and Gaziulusoy, Idil},
title = {Reinventing the Wheel: The Future Ripples Method for Activating Anticipatory Capacities in Innovation Teams},
year = {2022},
isbn = {9781450393584},
publisher = {Association for Computing Machinery},
address = {New York, NY, USA},
url = {https://doi.org/10.1145/3532106.3534570},
doi = {10.1145/3532106.3534570},
booktitle = {Proceedings of the 2022 ACM Designing Interactive Systems Conference},
pages = {387–399},
numpages = {13},
location = {Virtual Event, Australia},
series = {DIS '22}
}

@article{BENDOR2021,
title = {Looking backward to the future: On past-facing approaches to futuring},
journal = {Futures},
volume = {125},
pages = {102666},
year = {2021},
issn = {0016-3287},
doi = {https://doi.org/10.1016/j.futures.2020.102666},
url = {https://www.sciencedirect.com/science/article/pii/S0016328720301567},
author = {Roy Bendor and Elina Eriksson and Daniel Pargman}
}

@inproceedings{Alfrink_IASDR23,
    author = {de Kreek, M. and Alfrink, K. and de Waal, M. and Kortuem, G. and Turel, T. and Visser, B. and Samson, L.},
    title = {When 'doing ethics' meets public procurement of smart city technology – an Amsterdam case study},
    booktitle = {in De Sainz Molestina, D., Galluzzo, L., Rizzo, F., Spallazzo, D. (eds.), IASDR 2023: Life-Changing Design, 9-13 October, Milan, Italy. } ,
    year = {2023},
doi = {https://doi.org/10.21606/iasdr.2023.520}
}

@article{Rozendaal2024,
author = {Rozendaal, Marco C. and Vroon, Jered and Bleeker, Maaike},
title = {Enacting Human–Robot Encounters with Theater Professionals on a Mixed Reality Stage},
year = {2024},
issue_date = {March 2025},
publisher = {Association for Computing Machinery},
address = {New York, NY, USA},
volume = {14},
number = {1},
url = {https://doi.org/10.1145/3678186},
doi = {10.1145/3678186},
journal = {J. Hum.-Robot Interact.},
month = nov,
articleno = {1},
numpages = {25}
}



\appendix


\section{INTERVIEW GUIDE} \label{Appendix:interview_guide}

\begin{enumerate}

\item Warm up questions: role in past or present automation projects

\begin{itemize}
    \item Can you tell me a little bit about the work you do/did in \textit{[anonymous airport]} as part of automation projects?
    \begin{itemize}
        \item Which projects were you part of?
        \item What was your role in each of those projects?
        \item What is the stage of each project at the moment?
    \end{itemize}
    
\end{itemize}

\hfill

\item Future visioning, automation tests, and suppliers: rationale and workflows

\begin{itemize}
    \item In those automation projects, was the future vision already given, or was this something you were responsible for developing?
    
    \begin{itemize}
     \item How do you translate a future vision  into a concrete project you are responsible for? 
    \item Can you tell me about the process you follow?
      \end{itemize}
    \item How do you decide on what tests to conduct in automation projects?
    \begin{itemize}
    \item How do you shape/define the ``best case scenario''? 
    \item How do suppliers participate in defining such a ``best case scenario''?

    \end{itemize}
    \item How do you engage with suppliers, mainly regarding the steps that bring you from a request for information to a concrete joint project plan?

\end{itemize}

\hfill

\item Fall back situations, human operators, and their interventions

\begin{itemize}
    \item \textit{In such a high stakes environment, the failure of autonomous equipment could lead to enormous effects...} 

    \begin{itemize}
        \item How do you currently design fall back situations?
      \item When during innovation timelines are fall back situations considered?
       \end{itemize}
\end{itemize}

\hfill

\item Help in their practice:
discussing a ``potential tool'' 

\begin{itemize}
    \item What kind of tools do you currently use to support your practice in automation projects?
    \item Can you think about any challenges in automation projects you would need support with?
    \begin{itemize}
        \item How could a potential tool help you? (e.g., with the process of designing fall back situations?)
    \end{itemize}
    
\end{itemize}

\hfill

    \item Finally, before closing this interview, do you have any other thoughts you would like to share with me regarding your experience in automation projects?

\end{enumerate}

\onecolumn

\section{OVERVIEW OF AUTOMATION PROJECTS}\label{Appendix: automation projects}
\begin{table*}[h]
\centering
\caption{Overview of automation projects conducted in Amsterdam Airport Schiphol within the \textit{Autonomous Airside Operations 2050} program, including their description, scope, and timeline and current status.}
\label{tab:projects overview}
\small
\begin{tabularx}{\textwidth}{@{}>{\raggedright\arraybackslash}p{0.12\textwidth} >{\raggedright\arraybackslash}X >{\raggedright\arraybackslash}p{0.1\textwidth} >{\raggedright\arraybackslash}p{0.22\textwidth}@{}}
\toprule
\textbf{Project} & \textbf{Description} & \textbf{Scope} & \textbf{Timeline and Status}  \\
\midrule
Aircraft inspections & Exploring the potential value of (autonomous) drones  for aircraft inspections and studying the requirements for their implementation into airside traffic 
& Airside
& \textit{Ongoing: deployment phase, after having tested 5 use cases with the main contractors.}  \\

\addlinespace
Baggage handling robot &  Integrating collaborative robots for baggage lifting and sorting into baggage carousels to reduce operators' physical strain & Baggage halls
& \textit{On hold, after proof of technology was conducted and initial robots were implemented. }\\

\addlinespace
Baggage transportation vehicle & Studying Autonomous Vehicles' integration requirements within baggage halls, airside, and ramp environments. 
Later, building a concept of operations for buffering cold baggage in peak times, to increase peak performance  & Airside and baggage halls
& \textit{Ongoing. Multiple driving tests were conducted on site.} \\

\addlinespace
Bus & Studying autonomous buses' technological feasibility, use cases, and passenger experience at the airside. Later, achieving remote operations.  & Airside
& \textit{Ongoing: decision point, after having tested for more than 20 months on airside.} \\

\addlinespace
Cleaning robot & Integrating robots to assist with floor cleaning activities & Terminal
& \textit{Ongoing: deployment and scaling, after tests were conducted with the main contractor.} \\

\addlinespace
FOD removal & Creating an autonomous process for Foreign Object Debris (FOD) detection and removal at the aprons to contribute to a seamless inbound flow & Airside
& \textit{Ongoing: tests are being conducted to study the capabilities of the technology.} \\

\addlinespace
Ground power connection & Reducing human intervention in the connection of incoming aircraft to ground power to reduce the emissions of aircraft engines & Airside
& \textit{On hold, after a proof of technology was conducted.} \\

\addlinespace
Lawn-mowing & Integrating lawn-mowing robots into the maintenance of airside grass fields & Airside
& \textit{Ongoing: deployment, after various proof of technology tests were conducted} \\

\addlinespace
Passenger Boarding Bridge (PBB) & Connecting PBBs to incoming flights autonomously, including door detection and bridge movements & Airside
& \textit{Ongoing: scaling up, after having been deployed in various gates.}\\

\addlinespace
Snow fleet & Exploring the potential value of an autonomous fleet for the removal of snow in the airfield & Airside
& \textit{On hold: in tender} \\

\addlinespace
Wheelchair  & Providing Persons with Reduce Mobility (PRMs) with self-driving wheelchairs to move from security to their gate without human assistance & Terminal
& \textit{On hold, after having conducted feasibility tests}  \\
\bottomrule
\end{tabularx}
\end{table*}

\end{document}